\documentclass[10pt,aps,twocolumn,showpacs,superscriptaddress,floatfix,nofootinbib]{revtex4-2}
\usepackage[utf8]{inputenc}
\usepackage[T1]{fontenc}
\usepackage[english]{babel}
\usepackage{amsmath,amsfonts}
\usepackage[colorlinks=true]{hyperref}
\usepackage{graphicx}
\begin{document}
\title{Non-perturbative computation of thermal conductivity\\ based on  Path Integral Monte Carlo methods}
\author{Vladislav Efremkin}
\affiliation{Center for Advanced Systems Understanding, Helmholtz Zentrum Dresden-Rossendorf, D-02826 G\"orlitz, Germany}
\author{Stefano Mossa}
\affiliation{Universit\'e Grenoble Alpes, CEA, IRIG-MEM-LSim, 38054 Grenoble, France}
\author{Jean-Louis Barrat}
\affiliation{Universit\'e Grenoble Alpes, CNRS, LIPhy, 38000 Grenoble, France}
\author{Markus Holzmann}
\affiliation{Universit\'e Grenoble Alpes, CNRS, LPMMC, 38000 Grenoble, France}
\date{\today}
\begin{abstract}
The calculation of thermal conductivity in insulating solids at temperatures below the Debye temperature is problematic, due to the breakdown of classical and semi-classical approaches. In this work, we present a fully non-perturbative quantum methodology to compute thermal conductivity based on Path Integral Monte Carlo (PIMC) simulations combined with the Green-Kubo linear response theory.
The method is applied to rare gas solids modeled by a Lennard-Jones potential, paradigmatic systems where quantum effects strongly affect both thermodynamic and transport properties. From PIMC simulations, we obtain the temperature-dependent phonon frequencies, lifetimes, and specific heat.
From the imaginary time correlations of the energy current, we extract the thermal transport coefficients based on a physically motivated prior. 
We show that the experimentally observed increase of the thermal conductivity of argon and neon
at low temperatures cannot be explained within a  Peierls-Boltzmann framework using phonon line-widths at equilibrium. In contrast, a distinct transport lifetime emerges from the analysis of heat-current correlations. Our results demonstrate that quantum Monte Carlo methods provide a robust, non-perturbative framework to investigate heat transport in insulating solids, beyond the limits of classical molecular dynamics without relying on perturbative or semi-classical approximations.
\end{abstract}
\maketitle
\paragraph*{Introduction.--} A heat flux $\mathbf J$ occurs inside the bulk of a material in the presence of a temperature gradient, $\nabla T$. In the limit of small temperature gradients, Fourier's law,
\begin{equation}
{\mathbf J}= -\underline{\kappa} \nabla T, 
\label{eq:fourier}
\end{equation}
defines the heat conductivity tensor, $\underline{\kappa}$, in a given thermodynamic state, {\it e.g.}, temperature and pressure/density. Arguably, heat transport in solids, as described above, is conceptually  the simplest non-equilibrium transport phenomenon. It is also an active area of research in physics and materials science,
in particular in the context of thermoelectric phenomena or nanoscale energy conversion~\cite{Chen2005,Cahill2014,Tian2014-hu}. Accordingly, its theoretical description and numerical predictions have attracted enormous interest, both historically~\cite{Peierls1929,ashcroft1976solid,Srivastava2022} and in the more recent literature, employing increasingly sophisticated approaches~\cite{Baroni2018,Baroni2019,Simoncelli2019,Barbalinardo2020}. 

The theory of heat conduction in crystalline solids has been established by Peierls a century ago~\cite{Peierls1929}. First-principle calculations of the thermal conductivity for molecular systems described by a model Hamiltonian,
\begin{equation}
H=\sum_i \frac{p_i^2}{2m_i} + \sum_{i<j} v(r_{ij}),
\label{eq:hamiltonian}
\end{equation}
with $m_i$ the mass of particle $i$, have been mostly based either on Molecular Dynamics (MD) simulation or lattice dynamics within the Boltzmann transport equation (BTE). In recent years, important advances have been achieved in predictive first-principles calculations without adjustable parameters, overcoming limitations of the Peierls--Boltzmann approach, and extending their applicability to amorphous and partially disordered systems,
based on Wigner's dynamics or the quasi-harmonic Green-Kubo (QHGK) framework~\cite{Broido,Baroni2019,Simoncelli2019,Barbalinardo2020,Simoncelli2022,Caldarelli2022,Fiorentino2023}. 
While these approaches represent a significant progress providing a unified framework for crystalline and amorphous systems, 
and are now implemented in several open-source packages
\cite{ShengBTE,PhonTS,Phono3py,almaBTE,Alamode,kaldo}, 
they rely on perturbative treatments of anharmonicity, generally restricted to three-phonon scattering processes. It has been demonstrated, however, that fourth and higher order processes can lead to a significant reduction of the heat conductivity
~\cite{Feng2016,Feng2017}, and also modify phonon properties~\cite{Morresi2021,Castellano2023,Castellano2025}.

To date, fully non-perturbative computational methods have been  based on classical or semi-classical dynamics. Their reliability for temperatures significantly below the Debye temperature, $T_D$, is thus highly questionable, as they either completely fail to reproduce the expected drop of the specific heat, or incorporate this feature using {\it ad-hoc} corrections to the classical result. Although classical MD calculations for solid argon lead to surprisingly good agreement with experiment, down to low $T$, any attempt to empirically include quantum effects to describe the heat capacity has degraded the apparent agreement, and has also failed to reproduce the  sharp increase of the heat conductivity observed at low temperatures~\cite{JLB2014}. 

Here, we employ Path Integral Monte Carlo (PIMC) methods~\cite{Ceperley95} to overcome these limitations. We calculate the heat conductivity within the Green-Kubo linear response theory, based on a simple yet robust physical model underlying the spectral reconstruction of the imaginary time current-current correlations. Our approach fully accounts for anharmonic and quantum effects, without relying on any perturbative expansion. Focusing on Lennard-Jones (LJ) model systems with parameters calibrated for solid argon (Ar) as well as solid neon (Ne), we show that it  captures correctly the sharp increase of thermal conductivity on lowering temperature below $T_D$ in both materials. Note that our methodology does not rely on the presence of perfect crystalline structures, and can therefore be applied to any localized system, {\it e.g.}, disordered or amorphous solid. In particular, similar to the crystalline solids discussed below, accessible imaginary-time correlation functions also enable to directly test and verify spectral functions of the harmonic~\cite{Allen1989,Allen1993} or QHGK theory~\cite{Fiorentino2024} for the glass state, and naturally extend them into the quantum regime.

\paragraph*{Theory.--} Based on Green-Kubo linear response, we obtain the thermal conductivity at thermodynamic equilibrium from the energy-current correlation function as,
\begin{equation}
\kappa = \frac{1}{3}\operatorname{Tr}\underline{\kappa}
=\frac{k_B \beta^2}{3} \sum_\alpha 
\lim_{\omega \to 0} \Lambda_{\alpha \alpha} (\omega),
\label{eq:k_current}
\end{equation}
where $\beta=(k_B T)^{-1}$ is the inverse temperature, and
\begin{eqnarray}
\Lambda_{\alpha\alpha}(\omega) &=& \frac{1}{V}\lim_{\eta \to 0^+}\text{Re} \int_0^\infty dt e^{i(\omega+i\eta) t}
\langle J_\alpha(t)J_\alpha(0) \rangle \nonumber \\
&=& \frac{\pi}{V}\!\sum_{n,m}\!\frac{e^{-\beta E_n}}{Z}
|\langle E_n | J_\alpha |E_m\rangle |^2
\delta(\omega\!-\!E_{mn}), 
\label{eq:Lambda}
\end{eqnarray}
defines the spectral function. Here, $V$ is the system volume, $|E_n\rangle$ is an energy eigenstate of energy $E_n$, $Z=\operatorname{Tr} e^{-\beta H}$, $E_{mn}=E_m-E_n$, and the $J_\alpha$ are the $\alpha=x,y,z$ components of the heat current operator, $\mathbf{J}$, discussed below. Within numerical PIMC methods, we are able to compute imaginary time correlation functions controlling both systematic and statistical biases, as
\begin{eqnarray}
C_{\alpha \alpha}(\tau) &=& \frac1V
\langle J_\alpha(\tau) J_\alpha(0) \rangle
\label{Coftau} \\
&\equiv&
\frac{1}{ZV} \operatorname{Tr}
\left[ e^{-(\beta-\tau) H} J_\alpha e^{-\tau H} J_\alpha \right],\nonumber
\label{eq:c_tau_def}
\end{eqnarray}
which implicitly provides $\Lambda_{\alpha \alpha} (\omega)$ using
\begin{equation}
C_{\alpha \alpha}(\tau)	=\frac{1}{\pi}
\int_0^\infty d\omega\; \Lambda_{\alpha \alpha}(\omega)
\left[e^{-\tau \hbar \omega} + e^{-(\beta-\tau)\hbar \omega} \right],
\label{LambdaofOmega}
\end{equation}
as can be checked by explicit insertion of the respective spectral representations.

Inversion of Eq.~(\ref{LambdaofOmega}), which corresponds to an analytical continuation problem \cite{Jarrell1996,Sandvik2023}, is strongly affected by the stochastic error of $C_{\alpha\alpha}(\tau)$, intrinsic to any Monte Carlo method. A straightforward PIMC estimator for the  heat flux operator, see Eqs.~(\ref{Jharmonic}) and~(\ref{Jcurrent}), obtained by direct application of the momentum operator on the discretized path, suffers from large variance which diverges in the limit of vanishing imaginary time discretization, similar to the case of the direct kinetic energy estimator~\cite{Ceperley95}. We therefore use a variance-reduced {\it virial} estimator, as derived in~\cite{HO_PIMC,Carleo2013}. Spectral densities used as priors for the reconstruction of imaginary time correlations are based on physical considerations discussed later.

Dynamics at temperatures well below melting is usually dominated by normal mode vibrations (phonons in crystals), obtained from an an effective potential energy arising from deviations from the equilibrium  sites. In the harmonic approximation, the effective potential energy, $V_h$, is approximated by a quadratic form in the lattice displacements, ${\mathbf u}_i={\mathbf r}_i-{\mathbf R}_i$, 
\begin{equation}
V_h=\frac12 \sum_{i\alpha j \beta} u_{i\alpha} K_{i \alpha,j\beta}
u_{j\beta}.
\end{equation}
For classical systems, the effective force constants, $K_{i\alpha,j\beta}$, are given by the Hessian of the total potential energy, and the eigenmodes $\left\{e_n^{i\alpha}\right\}$ of the dynamical matrix, $\sum_{j\beta} K_{i\alpha,j\beta} e_n^{j\beta}=m_i\omega_n^2 e_n^{i\alpha}$, determine the phonon frequencies $\omega_n$. Anharmonic and quantum effects introduce an effective temperature and density dependence. In particular, quantum corrections reduce the modal heat capacity, $c_n=k_B (\hbar \omega_n/k_B T)^2 e^{\hbar \omega_n/k_B T}/(e^{\hbar \omega_n/k_BT}-1)^2$, compared to the classical values, and, within an effective harmonic theory, we have $C_V^h=\sum_n c_n$.

The effective phonon frequencies and lifetimes can be determined
from the spectra associated with the imaginary time correlation functions of the phonon modes,
\begin{equation}
\!\!\langle q_n(\tau)q_n(0)\rangle = \int_0^\infty \!\!\!\!\!\! d\omega \; S_n(\omega)\!\!
\left[e^{-\tau \hbar \omega}+e^{-(\beta-\tau)\hbar \omega} \right].
\label{eq:phonons-corr}
\end{equation}
Here, the $q_n$ are the normal mode coordinates which diagonalize the Hessian matrix of the Lennard-Jones interaction.
Assuming a Lorentzian distribution for the spectral function, $S_n(\omega)$, we determine the effective phonon frequencies, $\omega_n$, and inverse lifetimes, $\Gamma_n^{ph}$, from the reconstruction of the phonons imaginary time correlations.

Phonons in the harmonic approximation usually provide a reliable quantitative description of specific heat measurements, including the Debye $\propto T^3$ dependence at low $T$. In this regime, one expects the heat current to also be well described by its harmonic approximation~\cite{Hardy63},
\begin{equation}
\mathbf{J}^h = \frac12 \sum_{i\alpha j\beta} (\mathbf{R}_i-\mathbf{R}_j)
K_{i\alpha,j\beta} u_{i\alpha} p_{j\beta}/m_j,
\label{Jharmonic}
\end{equation}
with $p_{j\beta}$ the momenta. (See End Matter for the full current expression.) In the following we focus on the harmonic current-current correlation functions, 
\begin{equation}
C_{\alpha \alpha}^{h}(\tau)= \langle J_{\alpha}^h(\tau) J_{\alpha}^h(0) \rangle/V.
\label{eq:harmonic_current}
\end{equation}
For an ideal harmonic crystal, where we neglect any anharmonic effect in the imaginary time evolution, the corresponding spectral density for $\omega>0$ can be written as the sum of a singular part, $\Lambda_\alpha^s(\omega,\Gamma \to 0)$, which develops a delta function at zero frequency, and a regular part, $\Lambda_\alpha^r(\omega,\Gamma \to 0)$, which does not contribute for $\omega \to 0$, {\it i.e.},
\begin{equation}
\Lambda_{\alpha \alpha}^{h,0}(\omega)
= \Lambda_{\alpha}^{s}(\omega,\Gamma \to 0)
+ \Lambda_{\alpha}^{r}(\omega,\Gamma \to 0).
\label{eq:harmonic_spectral}
\end{equation}
Here, $\Gamma$ refers generically  to a damping mechanism of the harmonic modes, which vanishes in the harmonic limit~\cite{Fiorentino2024}, and is sometimes used as an adjustable parameter~\cite{Allen1989}. Introducing the decomposition of the harmonic current in normal modes, we obtain
\begin{eqnarray}
\Lambda_\alpha^s(\omega,\{\Gamma_{nm}\})\!\!\!
&=&\!\!\! \frac{1}{V} \sum_{nm} (\nu_{nm}^\alpha)^2 
\frac{\hbar \omega_n}{e^{\beta \omega_n}-1}
\frac{\hbar \omega_m}{1-e^{-\beta \omega_m}} \\
&\times &\!\!\! 
\frac{(\omega_n+\omega_m)^2}{4\omega_n\omega_m}
\frac{\Gamma_{nm}}
{(\omega-\omega_m+\omega_n)^2+\Gamma_{nm}^2}\nonumber \\
\Lambda_\alpha^r(\omega,\{\Gamma_{nm}\})\!\!\!
&=&\!\! \frac{1}{8V}\!\!\! \sum_{nm} (\nu_{nm}^\alpha)^2\!
\frac{\hbar \omega_n}{1-e^{-\beta \omega_n}}
\frac{\hbar \omega_m}{1-e^{-\beta \omega_m}}\\
&\times&\!\!\! 
\frac{(\omega_n-\omega_m)^2}{2\omega_n \omega_m}
\frac{\Gamma_{nm}}{(\omega-\omega_n-\omega_m)^2+\Gamma_{nm}^2},\nonumber
\label{eq::Lambda0}
\end{eqnarray}
with,
\begin{equation}	
\nu_{nm}^\alpha=\frac{1}{2 \sqrt{\omega_n \omega_m}}
\sum_{i\beta j \gamma}
e^{n}_{i\beta} K_{i\beta j\gamma} e^m_{j\gamma} (\mathbf{R}_{i\alpha}-\mathbf{R}_{j\alpha}).
\label{eq:nu}
\end{equation}
In the QHGK approximation for the thermal conductivity~\cite{Baroni2019}, the broadening $\Gamma_{nm}\simeq\Gamma_n^{ph}+\Gamma_m^{ph}$ is given by  the inverse phonon lifetimes, $\Gamma_n^{ph}$, determined numerically from the anharmonic (cubic) terms of the energy employing the Fermi's golden rule.  

In a perfect crystal, the phonon frequencies, $\omega_{\mathbf{k}a}$, are characterized by the lattice wave vector $\mathbf{k}$ and (longitudinal or transverse) polarization, $a$, and the matrix elements $\nu_{nm}^\alpha$ of Eq.~(\ref{eq:nu}) essentially reduce to the group velocity~\cite{Hardy63}, $v_{\mathbf{k}a}=\partial \omega_{\mathbf{k}a}/\partial{\mathbf{k}}$, which is diagonal in momentum space. We, therefore, recover the Peierls-Boltzmann form of the thermal conductivity in the relaxation-time approximation (RTA),
\begin{equation}
\!\!\!\kappa_{\text{PB-RTA}} = \frac{k_B}{V} \sum_{\mathbf{k}a} c_{\mathbf{k}a} 
v_{\mathbf{k}a}^2 \tau_{\mathbf{k}a}^{ph} \equiv
\overline{\tau^{ph}_{\text{PB-RTA}}} \sum_{\mathbf{k}a} c_{\mathbf{k}a}v^2_{\mathbf{k}a}.
\label{KappaPB}
\end{equation}
Here, $c_{\mathbf{k}a}$ is the modal specific heat, and the $\tau_{\mathbf{k}a}^{ph}$ are provided by the equilibrium phonon lifetimes, $\tau_{\mathbf{k}a}^{ph}=1/2\Gamma_{\mathbf{k}a}^{ph}$. For later reference, we also defined a weighted phonon lifetime as $\overline{\tau^{ph}_{\text{PB-RTA}}}$. In general, however, one expects vertex corrections beyond the quasi-harmonic Green-Kubo approach  \cite{Caldarelli2022}, leading to transport lifetimes, $\tau_{\mathbf{k}a}^{tr}$, which may differ from the bare phonon lifetimes (calculated at thermal equilibrium), even if the latter are computed in a non-perturbative manner. 
The goal of the PIMC calculations we describe below, is to  naturally take into account all quantum and anharmonic effects.
\begin{figure}[t]
\vspace{-0.5cm}
\centering
\includegraphics[width=0.5\textwidth]{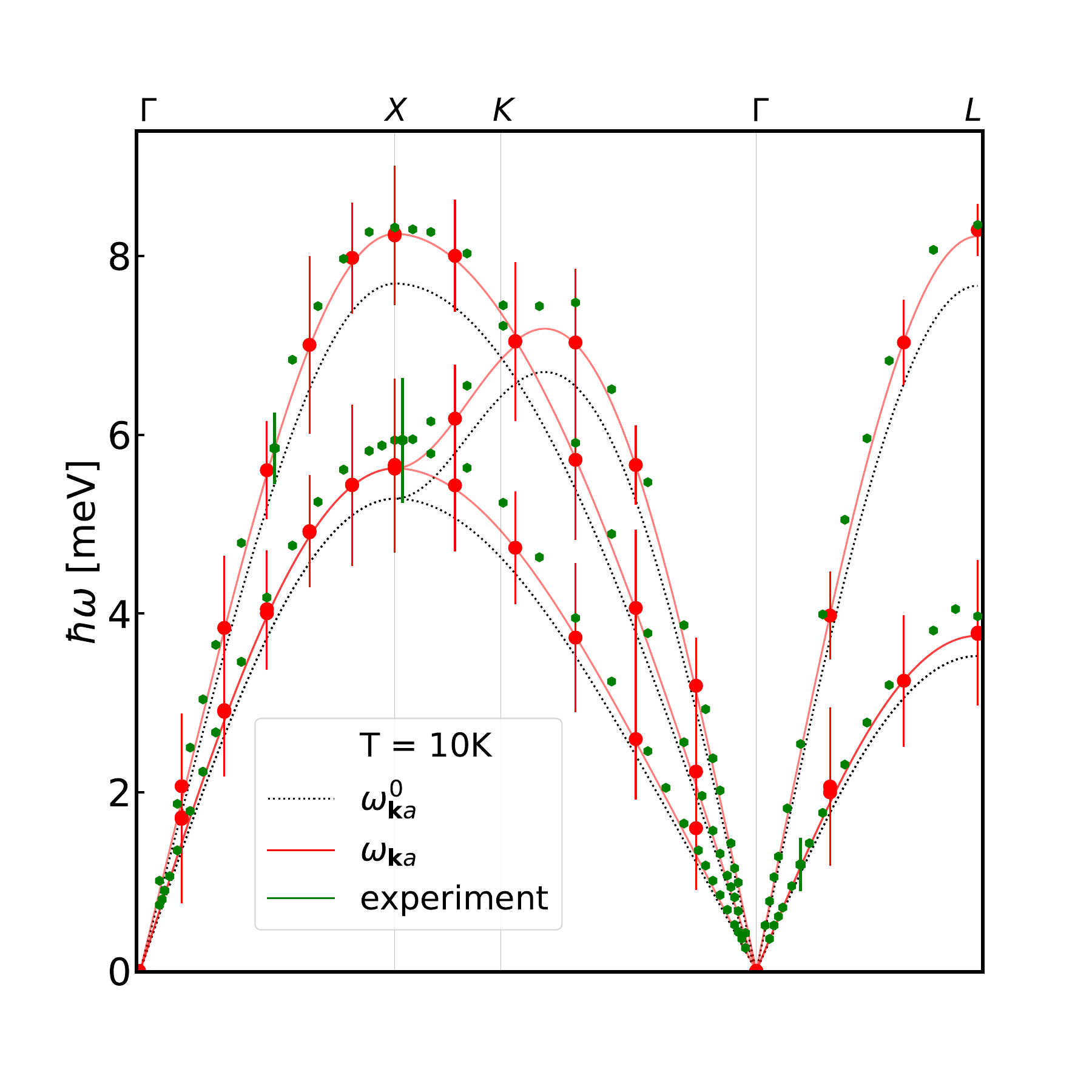}
\vspace{-1cm}
\caption{Phonon dispersions for solid argon, along the indicated directions of the wave-vector, $\mathbf{k}$. The results for longitudinal ($a=$~L) and transverse, ($a=$~T) phonons, $\omega_{\mathbf{k}a}$, (red circles) obtained from the normal modes correlation function for a simulated system with $N=$~864 atoms are compared with experimental data~\cite{Fujii1974}. The grey dotted lines indicate the harmonic frequencies, $\omega_{\mathbf{k}a}^0$, pertaining to the ideal harmonic crystal. The vertical (red) bars quantify the broadening of the phonon lines, $\Gamma_{\mathbf{k}a}^{ph}$, due to their finite lifetimes. The experimental line-widths~\cite{Fujii1974} are also indicated as vertical (green) bars, for selected data points.}
\label{Fig::om_disp}
\end{figure}

\begin{figure}[t]
\centering
\includegraphics[width=0.48\textwidth]{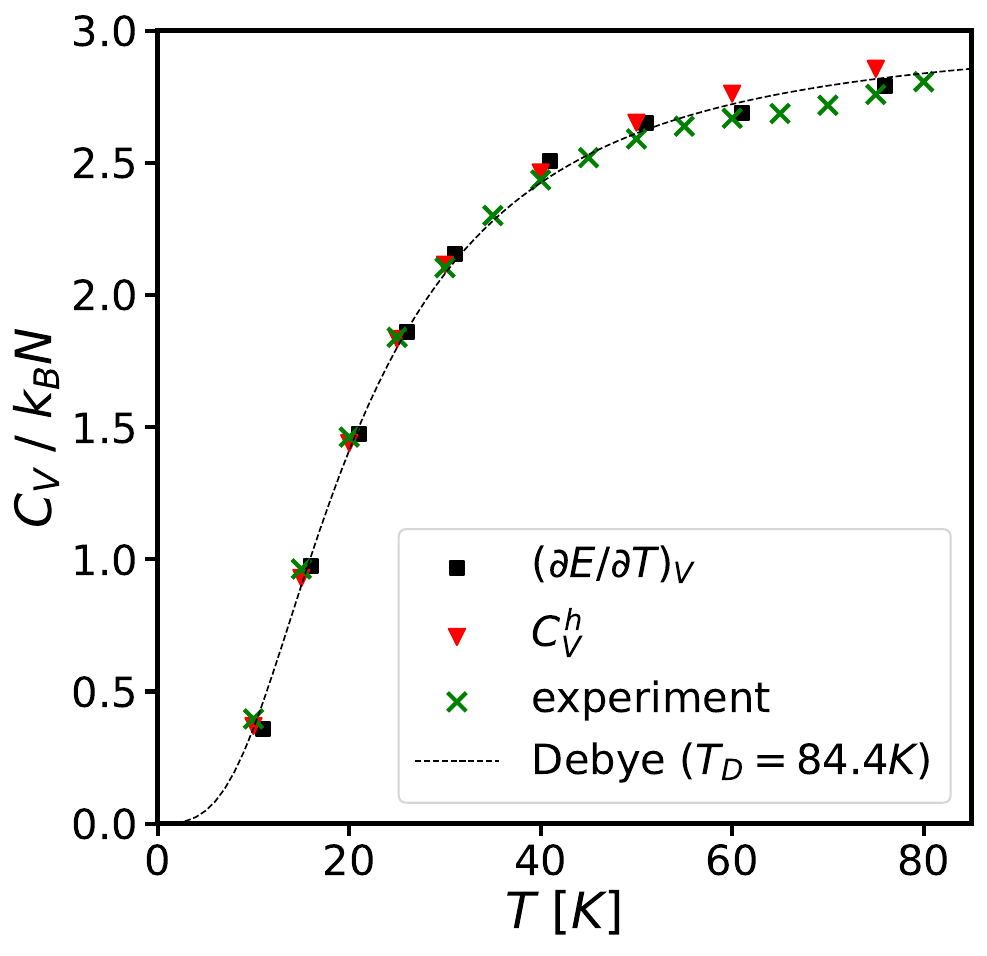}
\caption{Specific heat of solid argon from PIMC calculations compared to the experimental results. The black squares are obtained from the numerical derivative of the total energy, whereas $C_V^h$ (triangles) is the specific heat determined in the harmonic approximation based on the phonon frequencies, $\omega_{\mathbf{k}a}$, obtained from the normal mode correlation functions. The experimental data from~\cite{Peterson1966} ($\times$) are also shown for comparison, together with the corresponding values predicted by the Debye model with $T_D=$~84.4~K (dashed line).}
\label{Fig::specific_heat}
\end{figure}
\paragraph*{Results.--} We now apply the methodology outlined above to crystalline argon,  
modeled by a LJ potential, $v(r)=4 \epsilon \left[(r/\sigma)^{-12} - (r/\sigma)^{-6} \right]$, with $\epsilon=119.8$~K, $\sigma=3.405$~\AA  
~\cite{Cuccoli1993}, and a cutoff distance, $r_c=2.5 \sigma$. 
Unless explicitly stated, results 
are reported in LJ units, where time is expressed in units of $t_0 = 2.16\times 10^{-12}$~s, and the strength of quantum effects is encoded in the dimensionless parameter $Q=\sqrt{\hbar^2/m\sigma^2\epsilon}\simeq 0.0296$. Below $T\approx 75$~K, and close to ambient pressure, the atoms arrange in a fcc lattice. Our PIMC simulations are performed in the (NVT) ensemble, at densities $\rho=N/V$ varying from $\rho \sigma^3=1.052$ at $10$~K, to $0.98$ at $75$~K, approximately following 
the zero pressure isobar, with an imaginary time step discretization based on the primitive approximation~\cite{Ceperley95} corresponding to a temperature $\approx 800$~K. Below, we primarily present the results of computations with $N=108$, as we have not observed essential differences compared to larger systems in the temperature range considered. Note that here we do not particularly focus on a precise quantitative modelling of experiments, the case of argon serving as a first proof of concept. We have further applied the 
same methodology to calculate the thermal conductivity of solid neon, also shown below. The values of the parameters are provided in the supplementary material~\cite{supplementary}.

In Fig.~\ref{Fig::om_disp} we compare the  phonon dispersion resulting
from the normal mode correlations, Eq.~(\ref{eq:phonons-corr}), to the experiments of~\cite{Fujii1974}, at $T=10$~K. We observe an approximately $5\%$ shift compared to the bare (unperturbed) frequencies, $\omega_{\mathbf{k}a}^0$, computed from diagonalizing the Hessian of the LJ potential, which is in reasonable agreement with inelastic neutron scattering data. The obtained phonon lifetimes are also compatible with the experimental line-widths of~\cite{Fujii1974}.
\begin{figure}[t]
\centering
\includegraphics[width=0.48\textwidth]{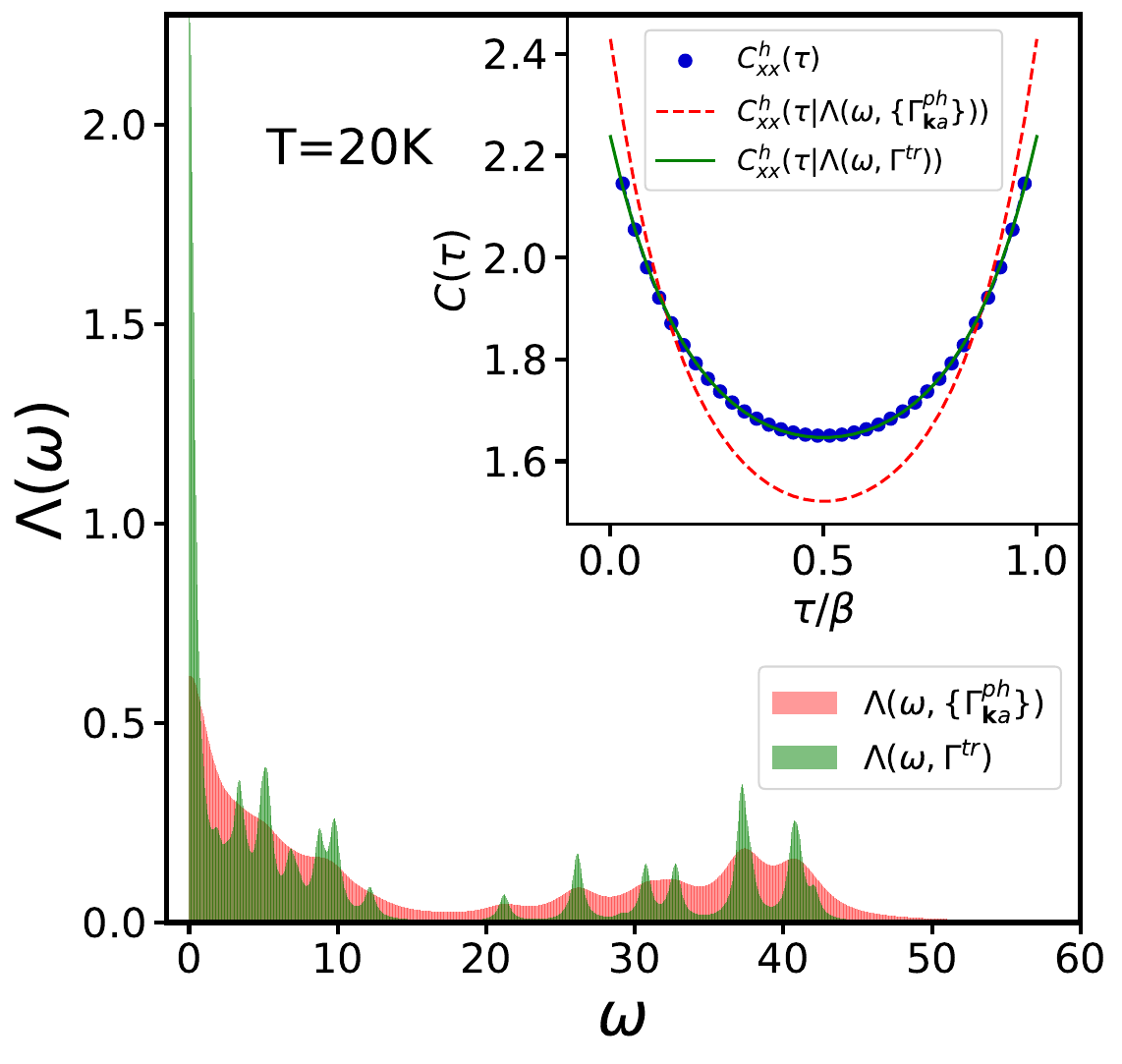}
\caption{
{\em Main panel:} Spectral functions
$\Lambda(\omega,\Gamma^{tr})$ (green) and $\Lambda(\omega,\{\Gamma^{ph}_{\mathbf{k}a}\})$ (orange)
of the harmonic current correlations at $T=20$K.
In the {\em Inset}, we show the corresponding  imaginary time correlations $C_{xx}^h(\tau |\Lambda(\omega,\Gamma^{tr}))$
and $C_{xx}^h(\tau|\Lambda(\omega,\{\Gamma^{ph}_{\mathbf{k}a}\}))$, as well as the
direct PIMC data $C_{xx}^h(\tau)$ (circles).
The best reconstruction obtained by imposing the inverse phonon lifetimes, $C_{xx}^h(\tau|\Lambda(\omega,\{\Gamma^{ph}_{\mathbf{k}a}\}))$ (dashed line), is incompatible with the calculated imaginary time correlation, and excludes the model spectra based
purely on phonon lifetimes.
}
\label{Fig::broad_fit_harm_sp}
\end{figure}

From our PIMC calculations we also have direct access to the temperature dependence of the total energy. In Fig.~\ref{Fig::specific_heat} we compare the resulting specific heat with experimental data~\cite{Peterson1966}, together with the harmonic approximation based on the phonon frequencies, $\omega_{\mathbf{k}a}$. 
An effective harmonic model based on the exact phonon frequencies provides a  quantitative description for the specific heat, systematically improving the predictions based on the bare frequencies obtained from diagonalizing the Hessian matrix of the interaction potential\cite{supplementary}.
Approaching the Debye temperature, deviations become visible, presumably due to stronger anharmonic effects which are not accurately grasped by an effective harmonic theory.

\begin{figure*}[t]
\centering
\includegraphics[width=\textwidth]{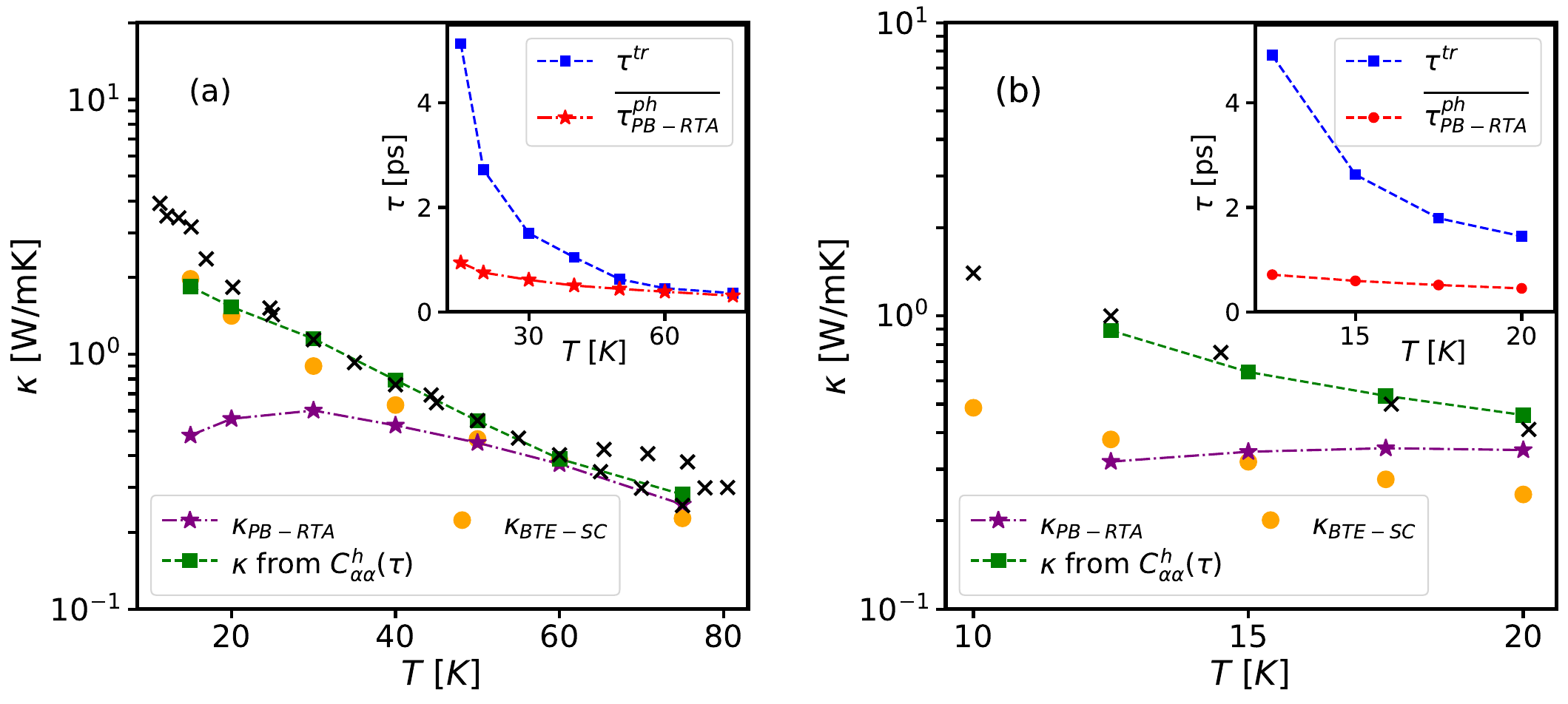}
\caption{
\textit{Main}: thermal conductivity of solid argon (a) and neon (b) versus temperature, obtained from the spectral reconstruction of the respective harmonic current correlation functions, $C_{xx}^h(\tau)$, compared with the result of the Peierls-Boltzmann equation, $\kappa_{\text{PB-RTA}}$ of Eq.~(\ref{KappaPB}), based on the PIMC phonon frequencies and lifetimes at equilibrium. The experimental data ($\times$), in black, are from Refs~\cite{christensen1975,krupski1969multiphonon,Weston1984}.
For comparison, we also show the results of self-consistent calculations based on Boltzmann transport equation,
$\kappa_{\text{BTE-SC}}$, as implemented in ~\cite{kaldo,Baroni2019}.
\textit{Inset:} 
transport lifetime, $\tau^{tr}$, resulting from the spectral reconstruction of the current correlations
in comparison to the weighted phonon lifetime, $\overline{\tau_{\text{PB-RTA}}^{ph}}$(see Eq.~(\ref{KappaPB})), for different temperatures. 
}
\label{Fig::kappa_current+phonon+md}
\end{figure*}

Once the equilibrium phonon lifetimes have been obtained from the normal-mode correlations, we can directly estimate
the heat conductivity from the Peierls-Boltzmann equation, Eq.~(\ref{KappaPB}). As shown in Fig.~\ref{Fig::kappa_current+phonon+md}, the determined values for $\kappa_{\text{PB-RTA}}$ qualitatively fail to reproduce the sharp increase of the thermal conductivity at temperatures below $T_D$. The behavior is actually similar to the results extracted from out-of equilibrium MD which heuristically include quantum effects for phonons to capture the decrease of the specific heat, see~\cite{JLB2014}. We have further verified that including off-diagonal elements of the velocity matrix, Eq.~(\ref{eq:nu}), as done in QHGK calculations, only introduce minor modifications. Our results for $\kappa_{\mathrm{PB\text{-}RTA}}$ therefore correspond to a QHGK calculation employing fully non-perturbative phonon lifetimes, rather than approximate lifetimes obtained from three- or four-phonon scattering rates within Fermi's golden rule~\cite{supplementary}.
Clearly, the disagreement with experiment  cannot be attributed to a failure of the determination of phonon frequencies and lifetimes, but instead points to a transport lifetime which becomes significantly different from $\tau_{\mathbf{k}a}^{ph}$ upon lowering $T$.

To move beyond the QHGK approximation and the related Peierls approximation, we now analyze the correlation functions of the harmonic current $C^h_{\alpha \alpha}(\tau)$ of Eq.~(\ref{eq:harmonic_current}), in order to reconstruct the spectral density. Instead of attempting a purely numerical inversion~\cite{Jarrell1996,Sandvik2023}, we maximally constrain the prior model based on the following physical considerations. First, from the spectral analysis of the phonon correlation functions we have determined the phonon frequencies, $\omega_{\mathbf{k}a}$, which provide an accurate description of the static energy fluctuations, as confirmed from the specific heat comparisons. As a consequence, it is justified to expect that the heat currents in the harmonic approximation will also provide an accurate description of thermal transport. 

Second, apart from broadening, the spectra of the harmonic current-current correlation function, should be close to the ideal harmonic spectra based on the extracted phonon frequencies. Note that the phonon frequencies already contain a temperature dependent anharmonicity shift of $\omega_{\mathbf{k}a}^0$. In order to take into account the broadening associated with anharmonicity, we expect the low frequency spectra to be well described by $\Lambda_\alpha^s(\omega,\Gamma^{tr})$, where $\Gamma^{tr}$ is a  mode independent broadening that may behave differently from the inverse phonon lifetimes. Since the detailed modifications of the regular part are beyond the scope of the present work, the prior  $\Lambda(\omega,\Gamma) \equiv \Lambda_\alpha^s(\omega,\Gamma)+\xi \Lambda_\alpha^{r}(\omega,\Gamma)$, using a mode independent broadening,  describes the  simplest model suitable for our purpose, as will be seen below. Indeed, we have found that the two parameters $\Gamma(T)$ and $\xi(T)$, are sufficient for a quantitative reconstruction of the imaginary time correlations in the entire temperature range, from $T=15$~K to melting~\cite{supplementary}.
 
In Fig.~\ref{Fig::broad_fit_harm_sp} we show the obtained spectral function, and the corresponding quantitative reconstruction of the imaginary time correlations. We further show that the corresponding one parameter model, $\Lambda(\omega,\{\Gamma_{\mathbf{k}a}^{ph}\})\equiv \Lambda_\alpha^s(\omega,\{ \Gamma^{ph}_{\mathbf{k}a}\})+\xi \Lambda_\alpha^{r}(\omega,\{\Gamma^{ph}_{\mathbf{k}a}\})$, imposing the exact mode resolved phonon lifetimes, fails to reconstruct the PIMC data for $C_{xx}^h(\tau)$.  This directly illustrates the sensitivity to $\Gamma$ of the imaginary time correlations, and, importantly, rejects the values of $\kappa_{\text{PB-RTA}}$ calculated with the full phonon lifetimes, entirely based on PIMC data without invoking the experimental results.

In Fig.~\ref{Fig::kappa_current+phonon+md} we show the resulting thermal conductivity obtained from our reconstruction for solid argon  as well as solid neon, using the same methodology. In both cases, our data closely follow the experimental values ($\times$), in particular they capture the rapid increase at lower temperatures, which is not reproduced by the use of equilibrium phonon lifetimes in the Peierls-Boltzmann equation. This qualitative failure of $\kappa_{\text{PB-RTA}}$ cannot be attributed to the neglect of the off-diagonal velocity matrix elements (see supplementary material~\cite{supplementary}).

Our reconstructed values of $\kappa$ depend on the choice of the prior, $\Lambda(\omega,T)$, particularly on the physical constraints built into the singular part, whereas they are robust against modifications of the regular part of the prior. Different model assumptions lead to variations of approximately $10\%$. The uncertainty in $\kappa$ is  dominated by this systematic error, which is significantly larger than the bias associated with the finite imaginary-time discretization. While we do not observe significant finite-size effects over the temperature range considered here, larger system sizes are expected to become necessary at lower temperatures in order to resolve the substantially longer phonon lifetimes. A detailed analysis of these sources of systematic uncertainty is provided in the Supplementary Material~\cite{supplementary}.

From the insets of Fig.~\ref{Fig::kappa_current+phonon+md},
 we finally see that the different behavior of $\kappa$ vs $\kappa_{\text{PB-RTA}}$ directly reflects 
 the strong increase of the extracted transport lifetime,
 $\tau^{tr}=1/2\Gamma^{tr}$, compared to the weighted phonon lifetime, $\overline{\tau^{ph}_{\text{PB-RTA}}}$.
Since our normal mode lifetimes
are not limited by perturbation theory, this difference hints to the necessity of including vertex corrections for the
current correlations at lower temperatures.

For comparison, results of self-consistent calculations of the Boltzmann transport equation
(BTE-SC), as implemented in ~\cite{kaldo,Baroni2019} are also shown in Fig.~\ref{Fig::kappa_current+phonon+md}; similar results from QHGK are given in the supplementary material \cite{supplementary}. Whereas BTE-SC (as well as QHGK) captures
the increase of $\kappa$ at low temperatures for solid argon, limitations of the underlying approximations are clearly visible for solid neon, whose strength of quantum effects,
$Q=\sqrt{\hbar^2/m \sigma^2 \epsilon}$, 
is roughly three times larger than for argon.

\paragraph*{Conclusions.--} In this work, we have introduced a general  methodology for calculating the thermal conductivity of insulating solids, making maximal use of the information extracted from equilibrium PIMC calculations. 
These provide, in a non-perturbative manner, the effective vibrational density of states, 
which constitutes the key ingredient for constructing a physically motivated
prior {\em ansatz} for the heat current correlation function. The latter is then employed to effectively invert the imaginary time correlation function, eventually giving access to the spectral density. Precise information on the singular structure of the spectral density is thus essential for our strategy to work.

Our calculations for solid argon and neon show that the method leads to results consistent with experiment, and that the rise of the thermal conductivity at low temperatures is basically due to a transport lifetime which differs from the equilibrium phonon lifetime obtained from normal mode analysis, as single phonon scattering does not lead to a full decorrelation of the heat current. This is illustrated in the inset of Fig.~\ref{Fig::kappa_current+phonon+md}, which shows that the transport lifetime increases sharply at low temperature (and presumably diverges as $T\rightarrow 0$ in a perfect crystal), while phonon lifetimes remain finite due to zero point motion.

Our method provides a computational framework for calculating heat transport in insulating solids within Path Integral Monte Carlo or Path Integral Molecular Dynamics methods. Although we have focused on a perfect, simple crystal, our approach directly extends to more complex crystals or amorphous systems, where it should provide a definitive assessment of the role and importance of quantum effects at low temperatures~\cite{Lory2017,Leggett2013,Zamponi2020}.

\paragraph*{Acknowledgements.--}
This work was supported by the project Heatflow (Grant No. ANR-18-CE30-0019) funded by the French national funding agency, Agence Nationale de la Recherche. We thank the Institut Laue Langevin, Grenoble, for hospitality.

\begin{figure}[t]
\centering
\includegraphics[width=0.48\textwidth]{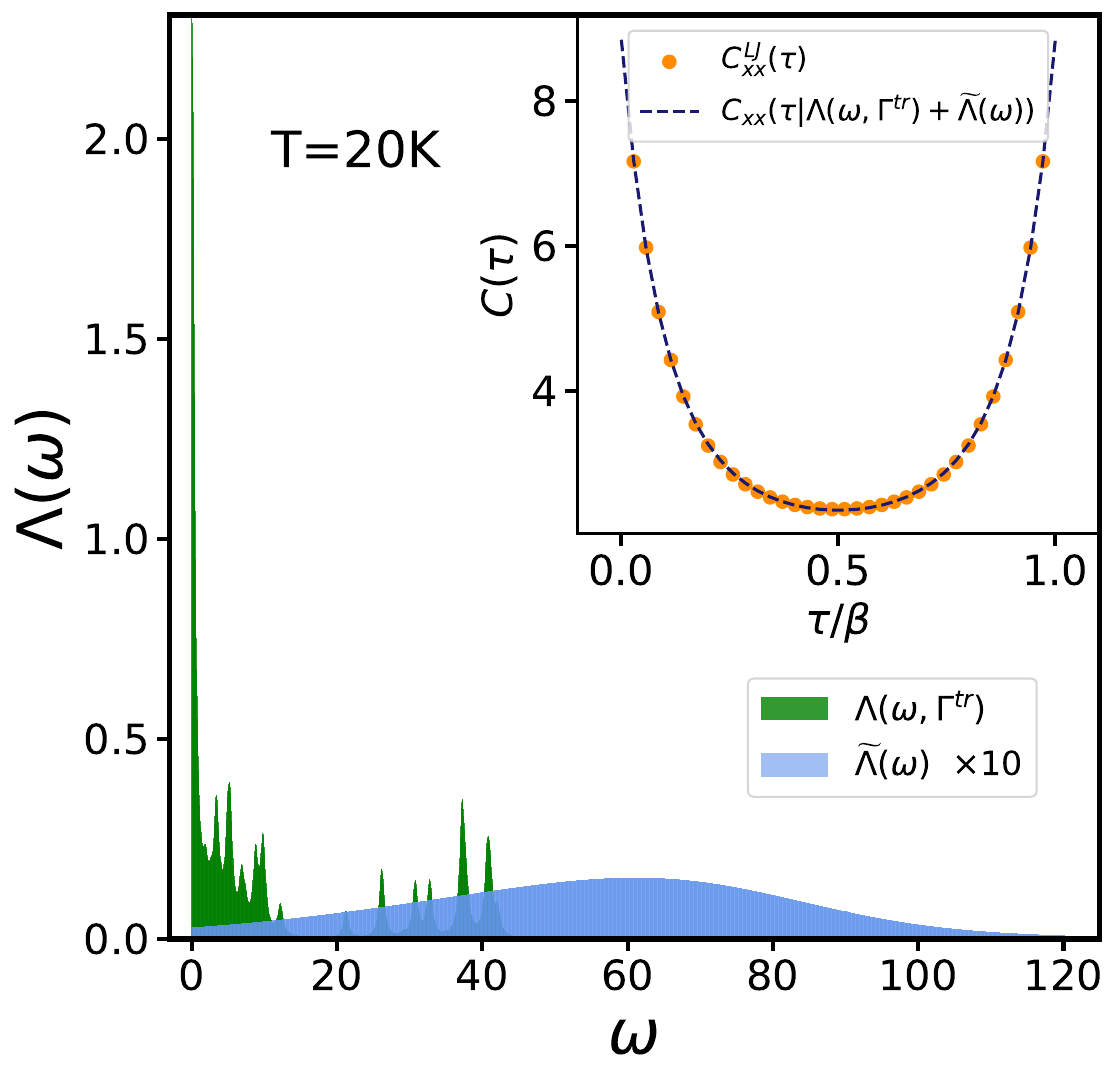}
\caption{Spectral function of total current current correlations at $T=20$K, based on the corresponding reconstruction
$\Lambda(\omega,\Gamma^{tr})$ of the harmonic current correlations
and the additional spectral weight $\widetilde{\Lambda}(\omega)$.
Notice the difference in scale between the two components of the spectral density.
The inset shows the reconstructed imaginary time correlation function of the total current obtained by optimizing only the 
additional weight $\widetilde{\Lambda}(\omega)$ 
without modifying $\Lambda(\omega,\Gamma^{tr})$.}
\label{Fig::totalJ}
\end{figure}
\vspace{-0.5cm}
\section*{End Matter}
\paragraph*{Total current.--} Within PIMC, it is straightforward to go beyond the harmonic approximation of the energy current and use the total energy current,
\begin{equation}
\mathbf{J} =
\frac12 \sum_{i \ne j} (\mathbf{R}_i-\mathbf{R}_j)  \left[
\mathbf{f}_{ij} \cdot (\mathbf{p}_i/m_i - \mathbf{u}) \right]
\label{Jcurrent}
\end{equation}
where $\mathbf{f}_{ij}=-\nabla_i v(r_{ij})$ and we have subtracted the center of mass velocity, $\mathbf{u}=\sum_i \mathbf{p}_i/(\sum_i m_i)$, to eliminate any spurious contributions due to the center of mass motion. We can then access and analyse the total current-current correlation function following exactly the same methodology as for the harmonic one.

However, in contrast to the harmonic current, the total current also contains higher order phonon transitions, affecting primarily the regular part of the spectral function, particularly visible in the correlations at short imaginary times. For a proper reconstruction, we have to include spectral weight, $\widetilde{\Lambda}(\omega)$, at higher frequencies, to  account for higher order transitions needed to for the reconstruction of the full current correlations in imaginary time. Since we do not expect these multi-phonon processes to  modify significantly the singular part already contained in the harmonic current expression, we can account for the difference by assuming a smooth density of states at the scale of typical phonon frequencies. Therefore, we can reconstruct the difference of the density of states, $\widetilde{\Lambda}(\omega)$,  by using conventional maximum entropy methods~\cite{Jarrell1996,Sandvik2023,HO_PIMC}.

In Fig.~\ref{Fig::totalJ} we show the obtained spectral reconstruction from the prior  $\Lambda(\omega,\Gamma^{tr})+\widetilde{\Lambda}(\omega)$. Here, $\widetilde{\Lambda}(\omega)$ is represented by a superposition of several Gaussians whose parameters are determined to reconstruct the imaginary time correlation function of the total current, whereas $\Lambda(\omega,\Gamma^{tr})$ is given by the unmodified reconstruction of the corresponding harmonic current correlations. As expected, the total current correlations contain higher frequency components, but with an order of magnitude smaller intensity. Even though entropy effects of this unconstrained reconstruction is likely smearing out sharp features, the additional spectral weight close to $\omega=0$ is too small to  modify significantly the value of the thermal conductivity. Within our precision, the thermal conductivity is quantitatively contained by the harmonic current correlations.
\bibliography{bibliography_lj_current.bib}
\widetext
\clearpage

\begin{center}
\textbf{\large Non-perturbative computation of thermal conductivity based on}
\end{center}
\vspace{-0.6cm}
\begin{center}
\textbf{\large Path-Integral Monte Carlo methods:}
\end{center}
\vspace{-0.5cm}
\begin{center}
\textbf{\large Supplementary Material}
\end{center}

\setcounter{equation}{0}
\setcounter{figure}{0}
\setcounter{table}{0}
\setcounter{page}{1}
\makeatletter
\renewcommand{\thetable}{S.\arabic{table}}
\renewcommand{\thefigure}{S.\arabic{figure}}
\renewcommand{\theequation}{S.\arabic{equation}}
\vspace{0.1cm}
\begin{center}
\large Vladislav Efremkin \\
\textit{\normalsize Center for Advanced Systems Understanding, Helmholtz Zentrum Dresden-Rossendorf, D-02826 G\"orlitz, Germany}
\end{center}
\begin{center}
\large{Stefano Mossa} \\
\textit{\normalsize Universit\'e Grenoble Alpes, CEA, IRIG-MEM-LSim, 38054 Grenoble, France}
\end{center}
\begin{center}
\large{Jean-Louis Barrat} \\
\textit{\normalsize Universit\'e Grenoble Alpes, CNRS, LIPhy, 38000 Grenoble, France}
\end{center}
\begin{center}
\large{Markus Holzmann} \\
\textit{\normalsize Universit\'e Grenoble Alpes, CNRS, LPMMC, 38000 Grenoble, France}
\end{center}
\vspace{0.5cm}
In the following, we present Supplementary Material (SM) discussing in detail the main sources of systematic error in the PIMC calculation of the thermal conductivity, together with additional results for solid neon and comparisons with various approximate methods.
\section{Imaginary time step discretization error}
\label{Sec:ImaginaryTime}

\begin{figure}[h]
\centering
\includegraphics[width=0.32\textwidth]{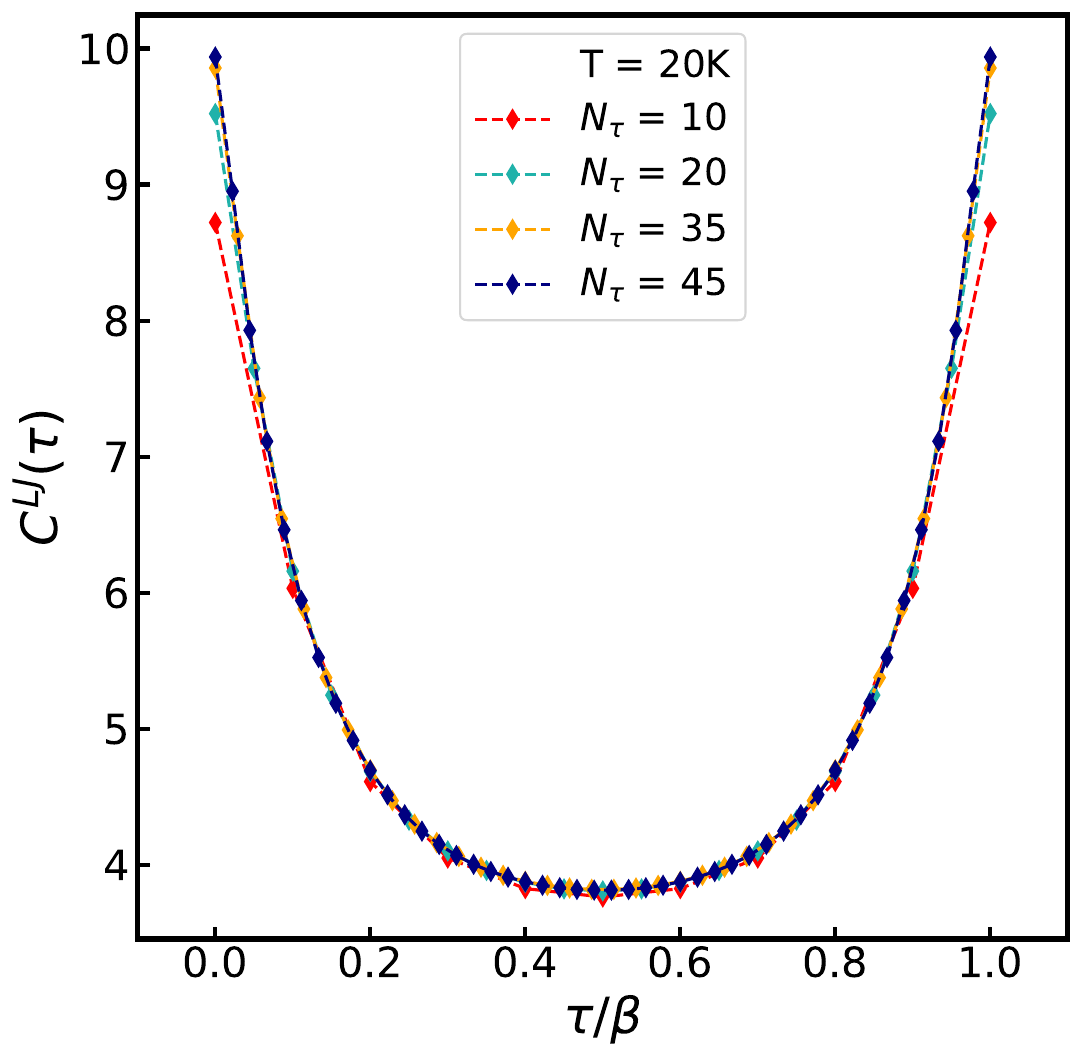}
\includegraphics[width=0.32\textwidth]{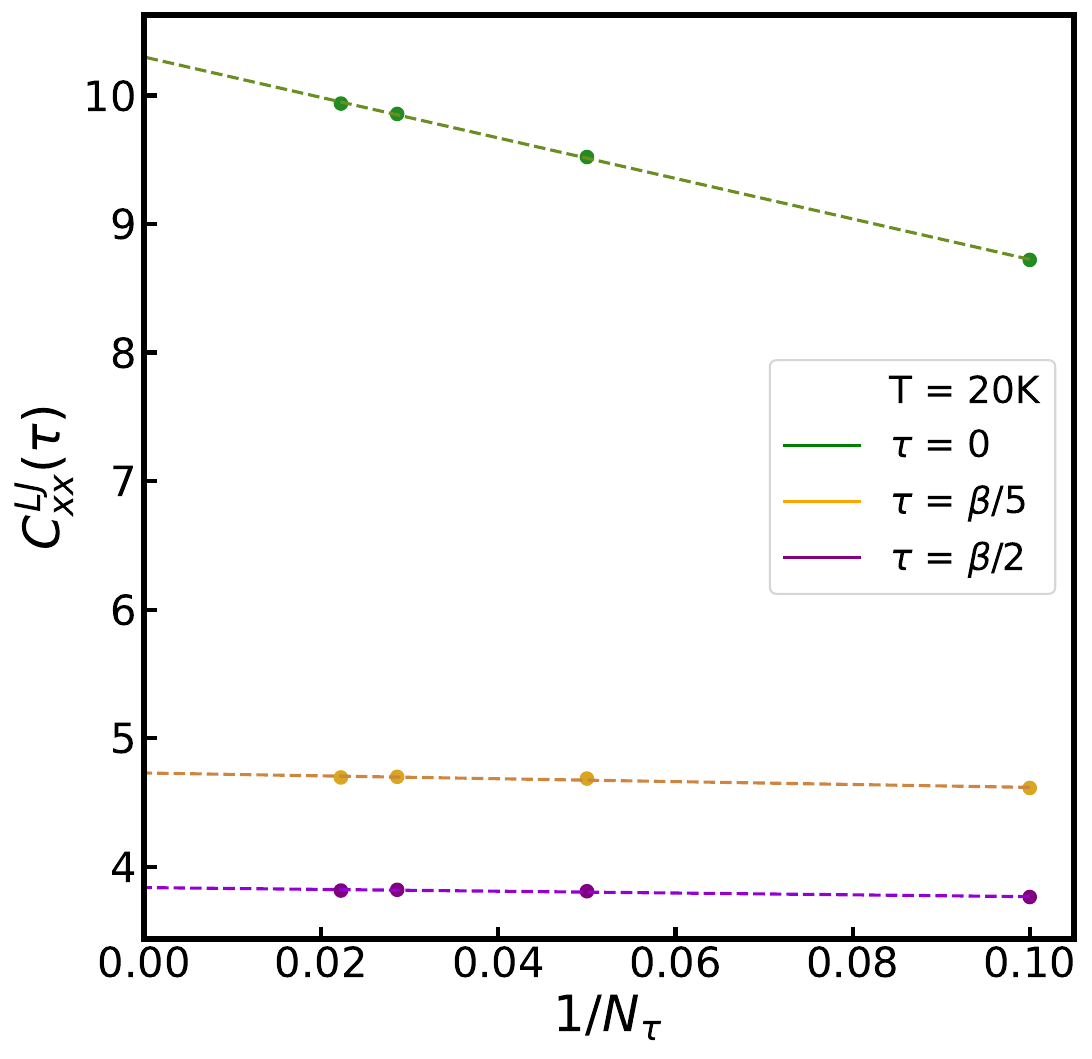}
\includegraphics[width=0.33\textwidth]{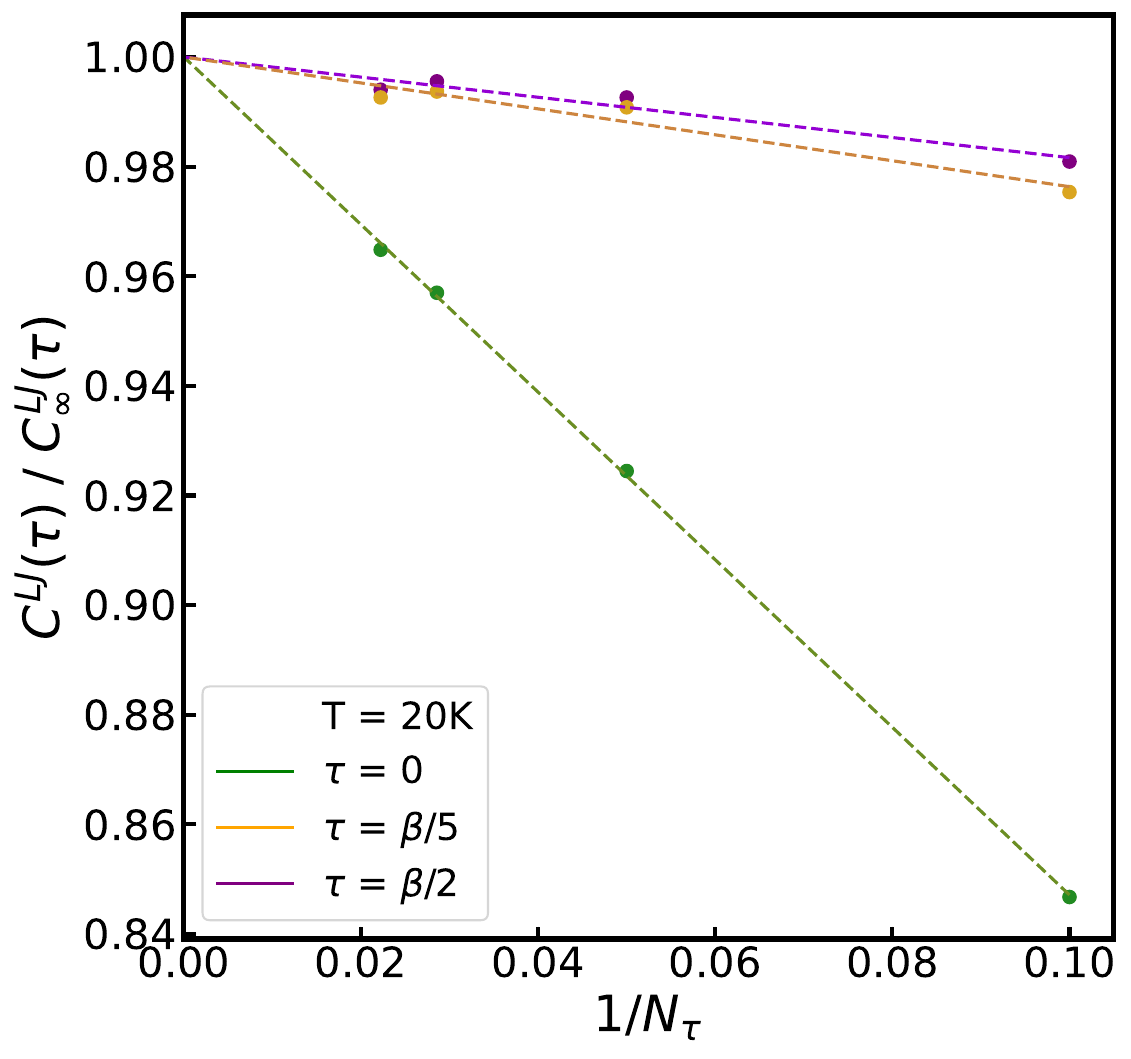}
\includegraphics[width=0.32\textwidth]{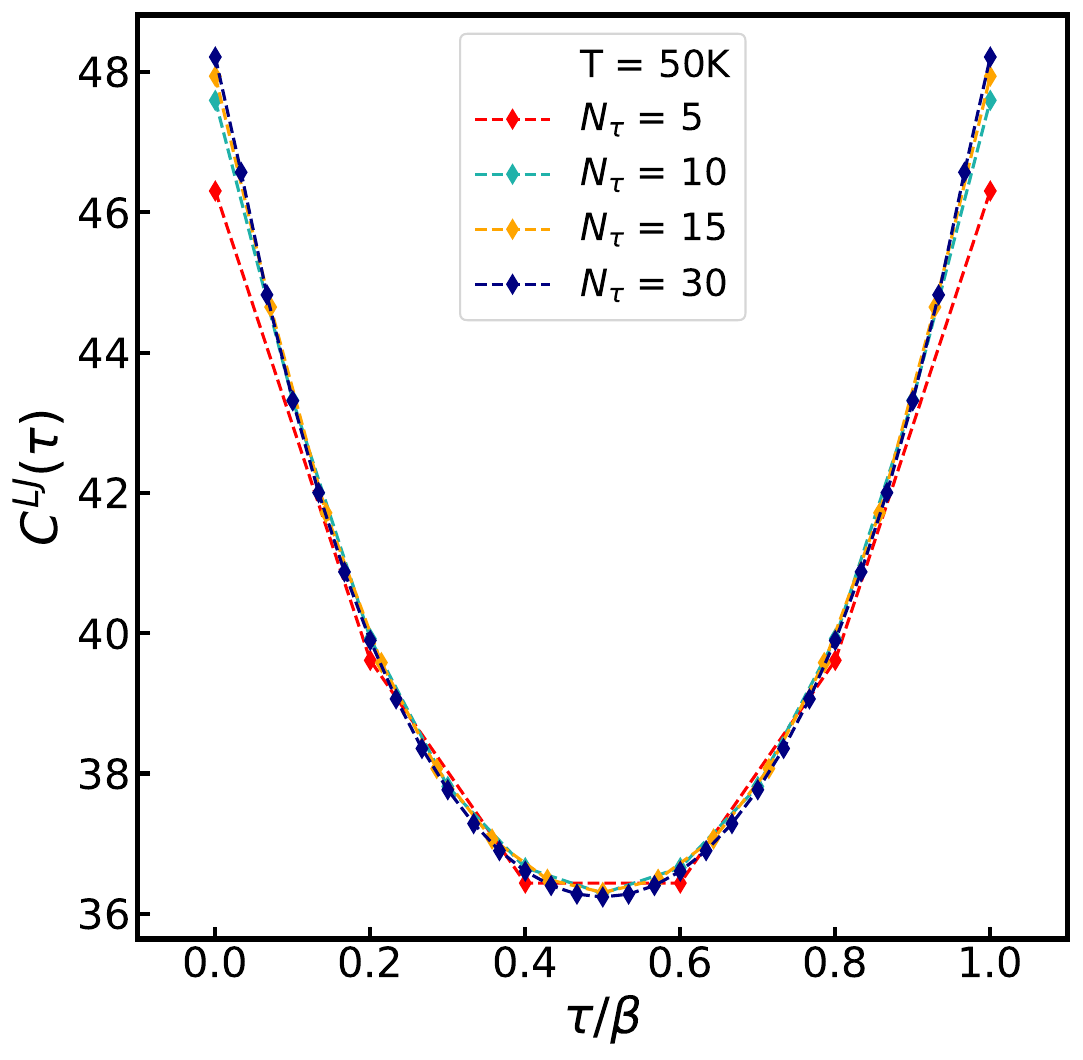}
\includegraphics[width=0.32\textwidth]{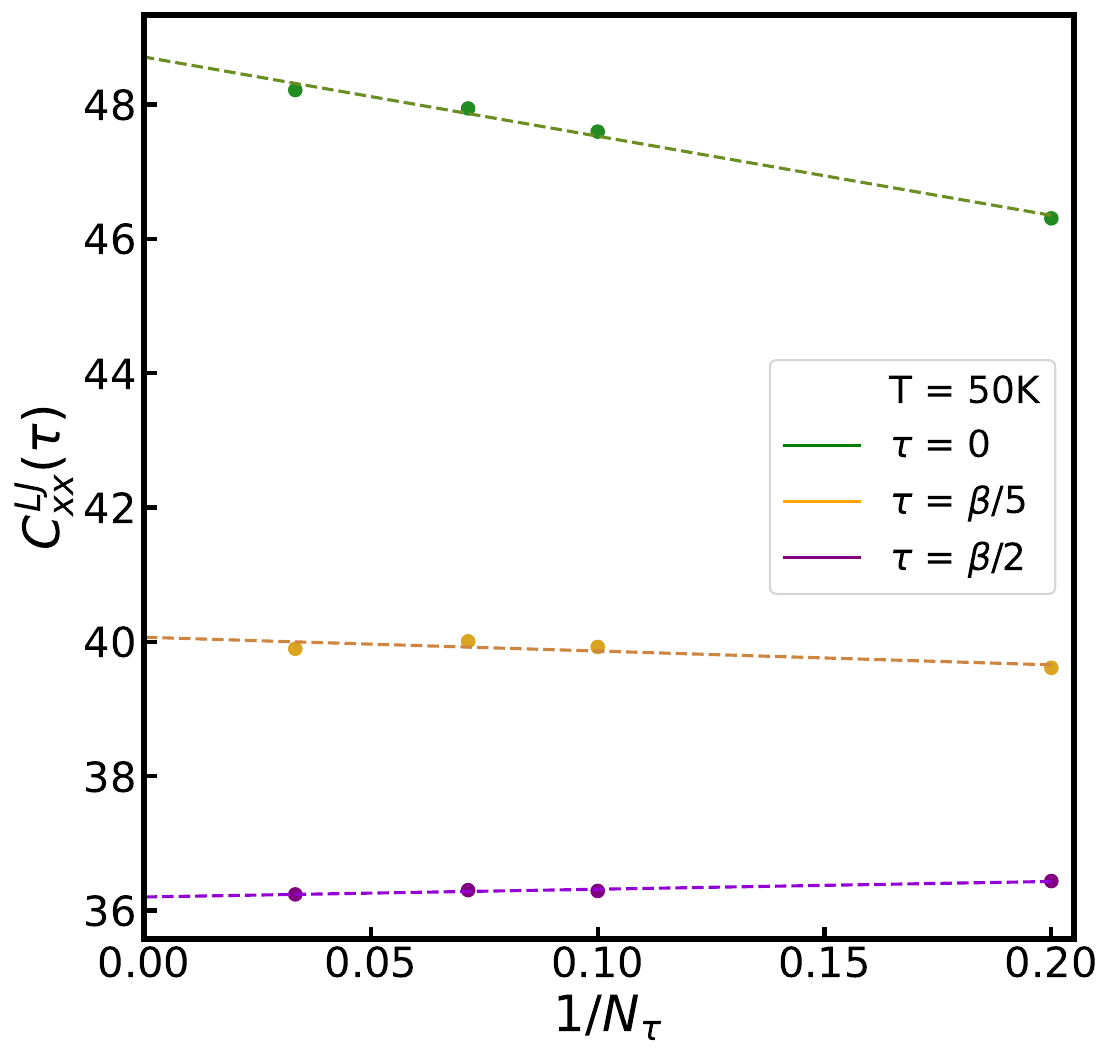}
\includegraphics[width=0.33\textwidth]{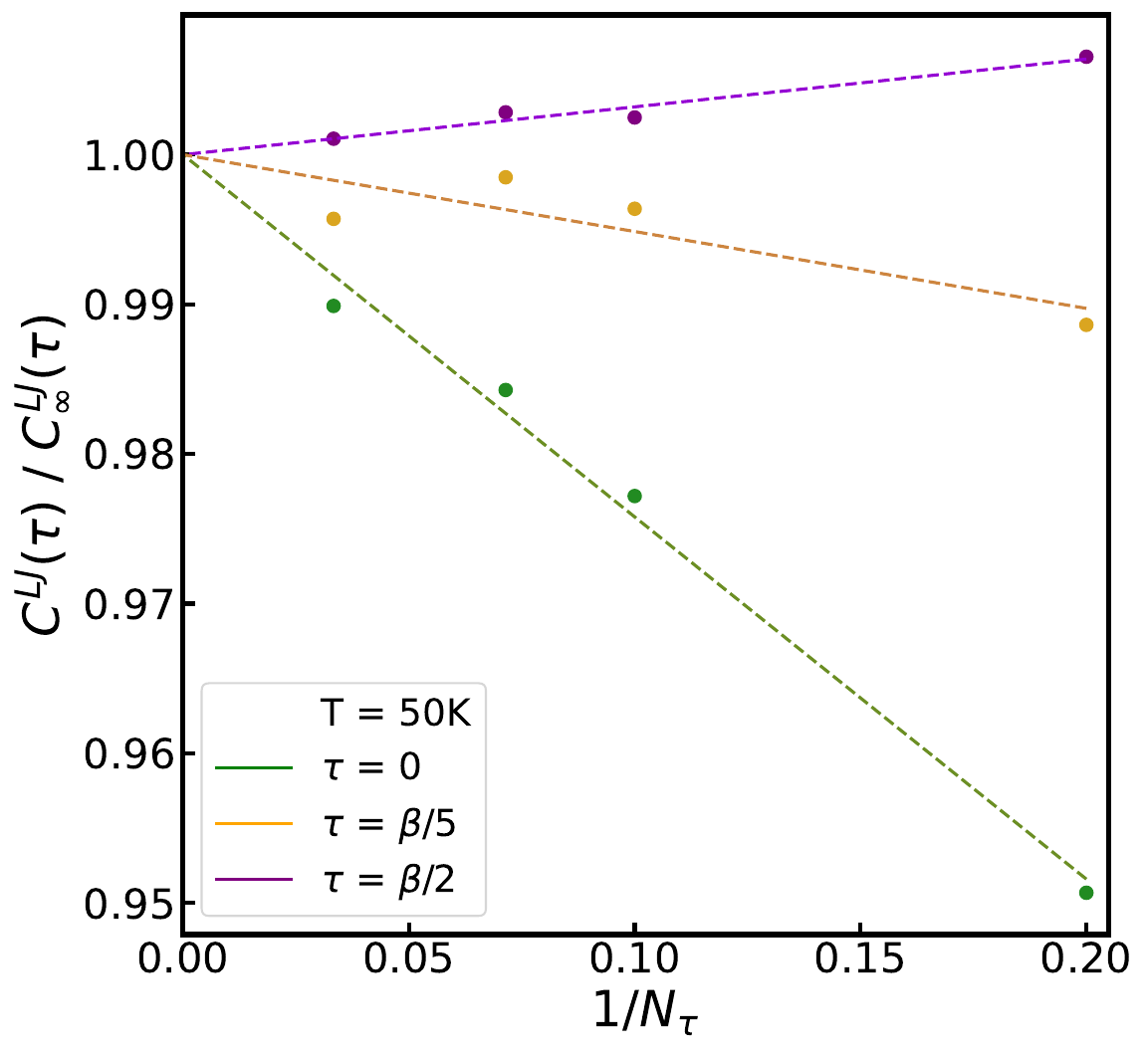}

	\caption{Study of the current-current correlation function as a function of imaginary times for $T=20$ and $50$K.
}
\label{Fig::tau}
\end{figure}

The temperature scale for quantum effects is given by the Debye temperature,
$84.4$K for Argon. The imaginary time step ( $\approx 800K$) 
 of the
simulations corresponds to approximately ten times higher 
effective temperatures, deep in the
classical regime.

More quantitatively, in Ref.~\cite{HO_PIMC} we have shown that the use of the primitive approximation at an inverse time $\tau$ 
to describe the density matrix of an harmonic oscillator of frequency $\omega$
merely introduces a shift of the harmonic frequencies of order $(\tau \hbar \omega)^2$ as long as $\tau \omega \ll 1$. Therefore, within the harmonic approximation, the discretization error due to a finite imaginary time step
$\tau$, will simply shift the phonon frequencies by the same order. These shifts add up to produce an overall first order shift of the effective temperature.  

The largest phonon frequency of the Lennard Jones potential in the
harmonic approximation is given by $\omega_\text{max} t_0 \approx 20$,
with $t_0=\sqrt{m \sigma^2/\epsilon}$. Then, $\hbar \omega_\text{max}
\approx 20 Q \epsilon$ is around $70$K, close to the
Debye temperature. For our choice of the time step,
we therefore expect only small shifts of the phonon frequencies of the order of percent. 
Alternatively, 
we can interpret these shifts as exact results due to a small (unknown)
modification of the underlying interaction potential.
Simulations performed at fixed imaginary time steps then
correspond to the use of this effective
potential which slightly deviates from the exact
Lennard Jones interaction. With respect to modelling 
solid Argon experiments,
these deviations are likely within the limit of the description
based on the Lennard-Jones pair potential to approximate the true
inter-atomic force fields.

\begin{figure}[h]
\centering
\includegraphics[width=0.7\textwidth]{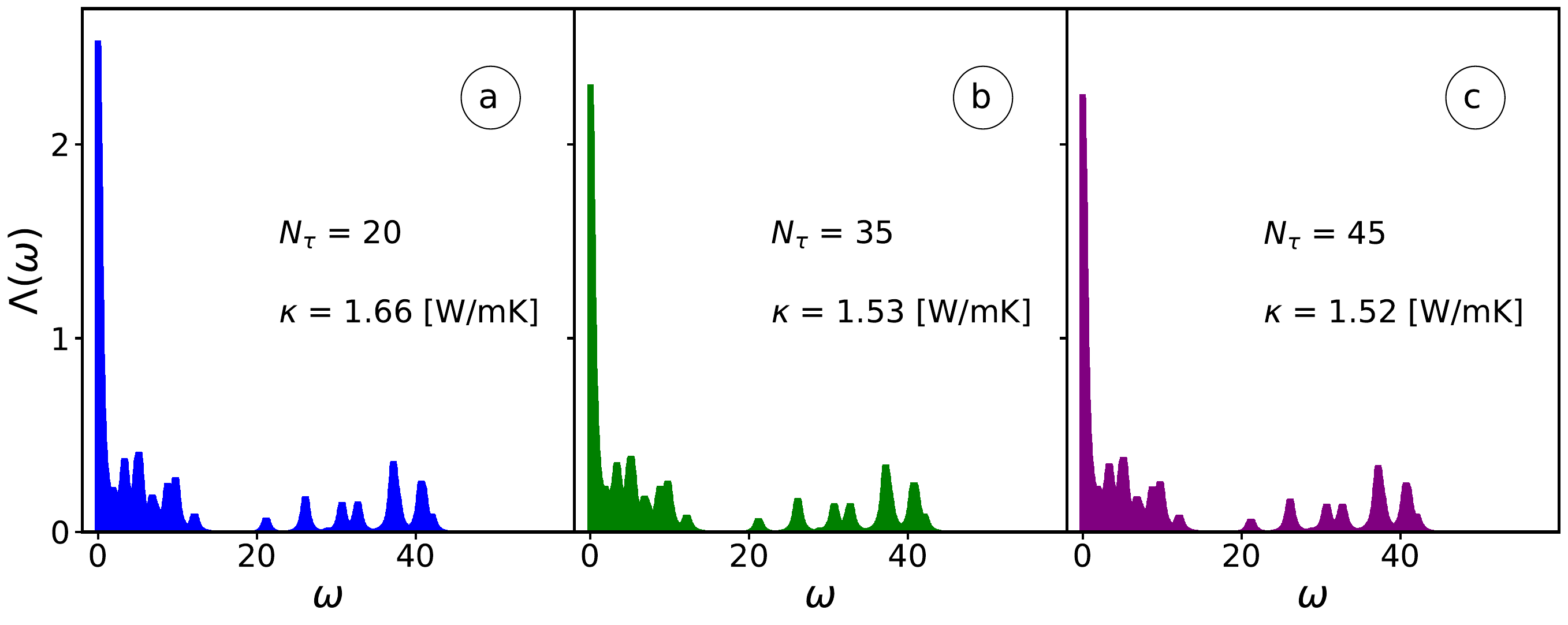}
\caption{Spectral functions of the heat current correlation functions of Ar at $T=20K$ reconstructed form different
imaginary time discretizations, corresponding to $N_{\tau}=20$, $35$, and $45$. The resulting values of heat conductance $\kappa$ obtained from the reconstructions are also shown.
}
\label{Fig::cur_time_step20}
\end{figure}

In order to quantify the influence of the finite imaginary time discretization step beyond the harmonic description on our quantities of interest,
we have explicitly studied the convergence of the current-current correlation
function in Fig.~\ref{Fig::tau} at $T=20$K and $50$K. We see, that
the time step discretization errors are most important
in the description of the correlations at vanishing imaginary time differences. At small, non-vanishing imaginary time differences, the error is considerably less, e.g. for $\tau=\beta/5$, the statistical
error is comparable to the systematic error for the chosen time step simulation. 
Since the heat conductivity is determined by the behavior of 
the correlations at large imaginary times,
around $\beta/2$, we expect that such a systematic error will not strongly influence the 
heat conductivity.
Indeed, the spectral functions obtained 
from the different imaginary time discretization, shown in Fig.~\ref{Fig::cur_time_step20}, and the resulting thermal conductivity
shows only small variations. Systematic effects due to the imaginary time discretizations should therefore not affect 
our results shown in the main text, corresponding to $N_\tau=35$.

\section{Finite Size Effects}

PIMC simulations are performed in simulation cells with periodic boundary conditions to model bulk materials, 
containing several 
crystalline unit cells. 
Despite the use of periodic boundary conditions, the finite number of atoms,
$N$, contained in the simulation cell, introduces a finite size bias with
respect to the thermodynamic limit, $N \to \infty$. 

Finite size errors can be related and estimated by analytical considerations,
similar to those frequently employed for electronic simulations at zero temperature \cite{FiniteN}. 
Such a detailed investigation goes beyond the
scope of the paper. Here, we estimate possible
finite size bias numerically, by comparing the outcomes of simulations
performed at different sizes. 

Let us note, that the spectral functions are dense around the origin
compared to the thermal energy $T$, already for $N=108$,
as can be seen
from the corresponding spectral function of the harmonic and total current, Figure 3 and 5 of the main text, respectively.
Therefore, we may expect that already our simulation cell of $N=108$ is large
enough to capture the thermal conductivity of the bulk up to smooth finite
size errors, presumably minor. The physical reason behind is that all
phonons, weighted by appropriate
matrix elements, contribute to the peak at $\omega=0$ of the current-current spectral function. Thus, the discretization of the phonons in the Brillouin zone
in the finite simulation cell
-- one major source of finite size effects, in general, --
is smoothed, and, therefore, also the thermodynamic limit extrapolation can be expected to be sufficiently regular.

\begin{figure}[h]
\centering
\includegraphics[width=0.48\textwidth]{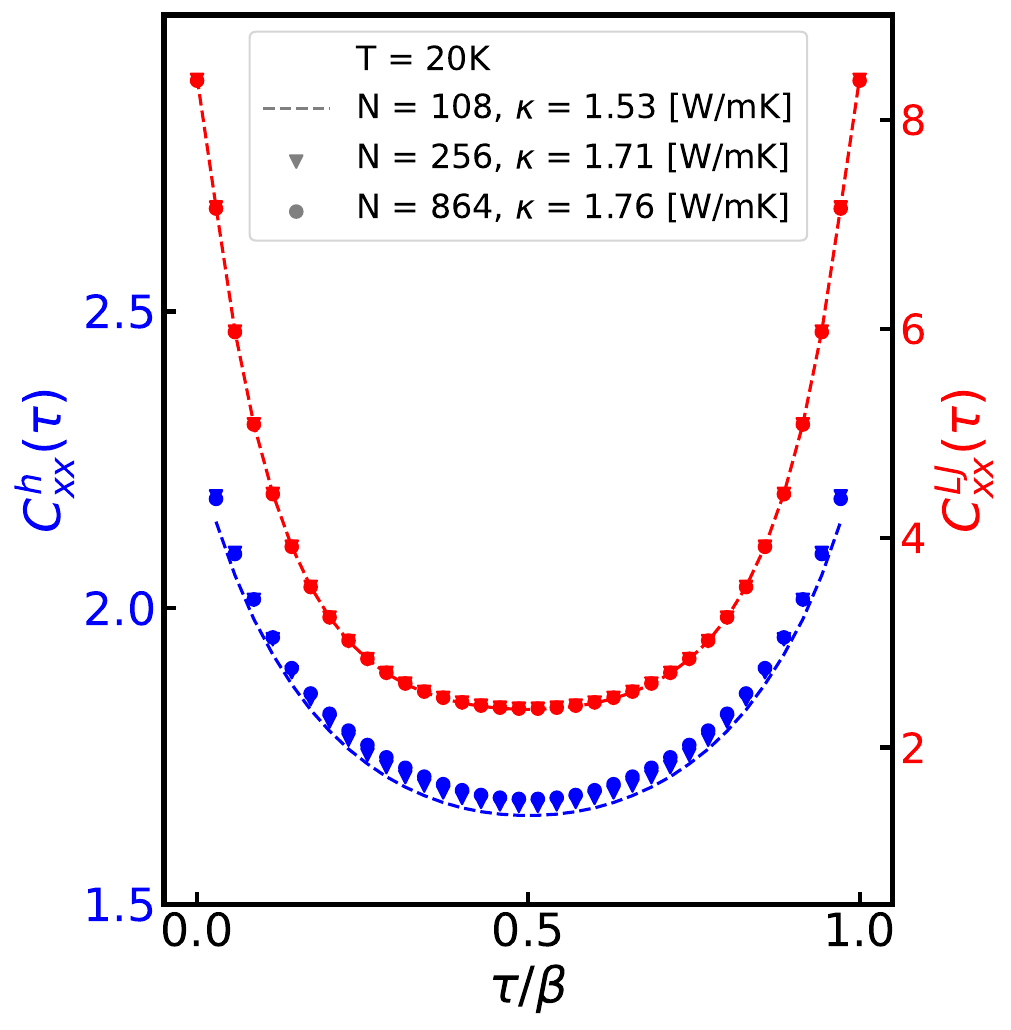}
\caption{Heat current correlation functions of Ar at $T=20K$ for systems of different number of atoms, $N=108$, $256$ and $864$. The upper (red) curves show the correlations of the full current, the lower (blue) lines the correlations of the harmonic current. The symbols of different system sizes are practically overlapping within the error. The corresponding values of the heat conductivity resulting from the spectral reconstruction is indicated in the labels.}
\label{Fig::cur_sizes}
\end{figure}

In Fig.~\ref{Fig::cur_sizes}, we show the imaginary time correlation function of the currents at $T=20$K for systems with $N=108$, $256$, and $864$ atoms, using the
same fixed cut-off in the LJ interaction parameter.
We have chosen $T=20$K as we expect finite size effects to be largest on the low temperature side.
Differences of the imaginary time correlation functions of different sizes are actually within the statistical accuracy for the correlations of the full currents, but
the harmonic current correlations show small differences
between $N=108$ and $N=256$. The resulting values of heat conductivity are noted in the same figure, and may indicate a finite size bias of $10-20\%$ of our 
$N=108$ particle simulation at $T=20$K. 
We will see below, that this is within the
uncertainty of the reconstruction of the spectral function from the imaginary time data.

\section{Reconstruction of spectral functions}
Within PIMC, we have access to the imaginary time correlation function of the current-current correlation function, 
$C_{\alpha \alpha}(\tau)$, essentially
a Laplace transform of the spectral function
$\Lambda_{\alpha \alpha}(\omega)$, we are interested 
in (see Eq.(6) of the main text). The numerical inversion is 
a notoriously difficult problem. This is in particular the case for the zero frequency limit of the spectral function: Within the finite stochastic error, different distributions of the spectral weight around $\omega=0$ will contribute essentially the same
to the signal of $C_{\alpha \alpha}(\tau)$, but does not 
sufficiently constrain the precise value of $\Lambda_{\alpha \alpha}$ at $\omega=0$.

To circumvent this problem, our reconstruction is based on
a physical model which constrains the spectral function by a minimal set of parameters to determine. 
In the following, we provide further indications to support the choice of our model function, and the robustness of the results
concerning possible variations of the parameters, focusing on the
harmonic current-current correlation function.

\subsection{Harmonic current correlations: Robustness of model assumptions}
\begin{figure}[h]
\centering
\includegraphics[width=\textwidth]{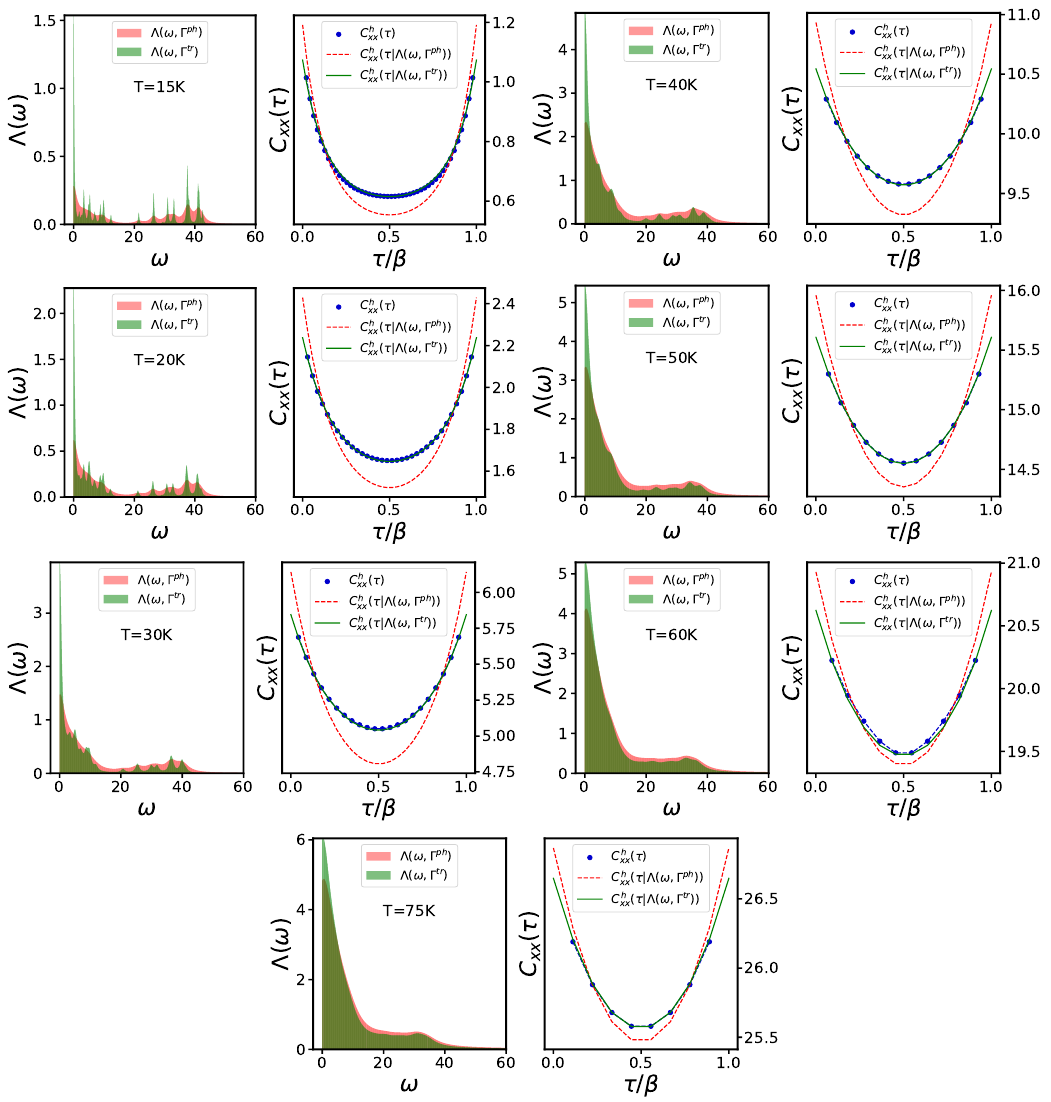}
	\caption{The harmonic current-current correlation functions in imaginary time $C_{xx}^h(\tau)$ at different temperatures, obtained from PIMC together with the reconstructed ones, $C_{xx}^h(\tau |\Lambda(\omega,\Gamma^{tr}))$ and $C_{xx}^h(\tau |\Lambda(\omega,\Gamma^{ph})$, as shown for a single temperature in Figure 3 of the main text. The default reconstruction based on the prior $\Lambda(\omega,\Gamma^{tr})$ determining the inverse transport lifetime $\Gamma^{tr}$ is compared to the best reconstruction obtained by $\Lambda(\omega,\Gamma^{ph})$ imposing the complete set of inverse phonon lifetimes, $\{ \Gamma^{ph}_{\mathbf{k}a} \}$. The resulting reconstruction imposing the phonon lifetimes is largely incompatible with the PIMC data of the imaginary time correlation at all temperatures.}
\label{Fig::sup_harm_T15_75}
\end{figure}
\begin{figure}[h]
\centering
\includegraphics[width=0.7\textwidth]{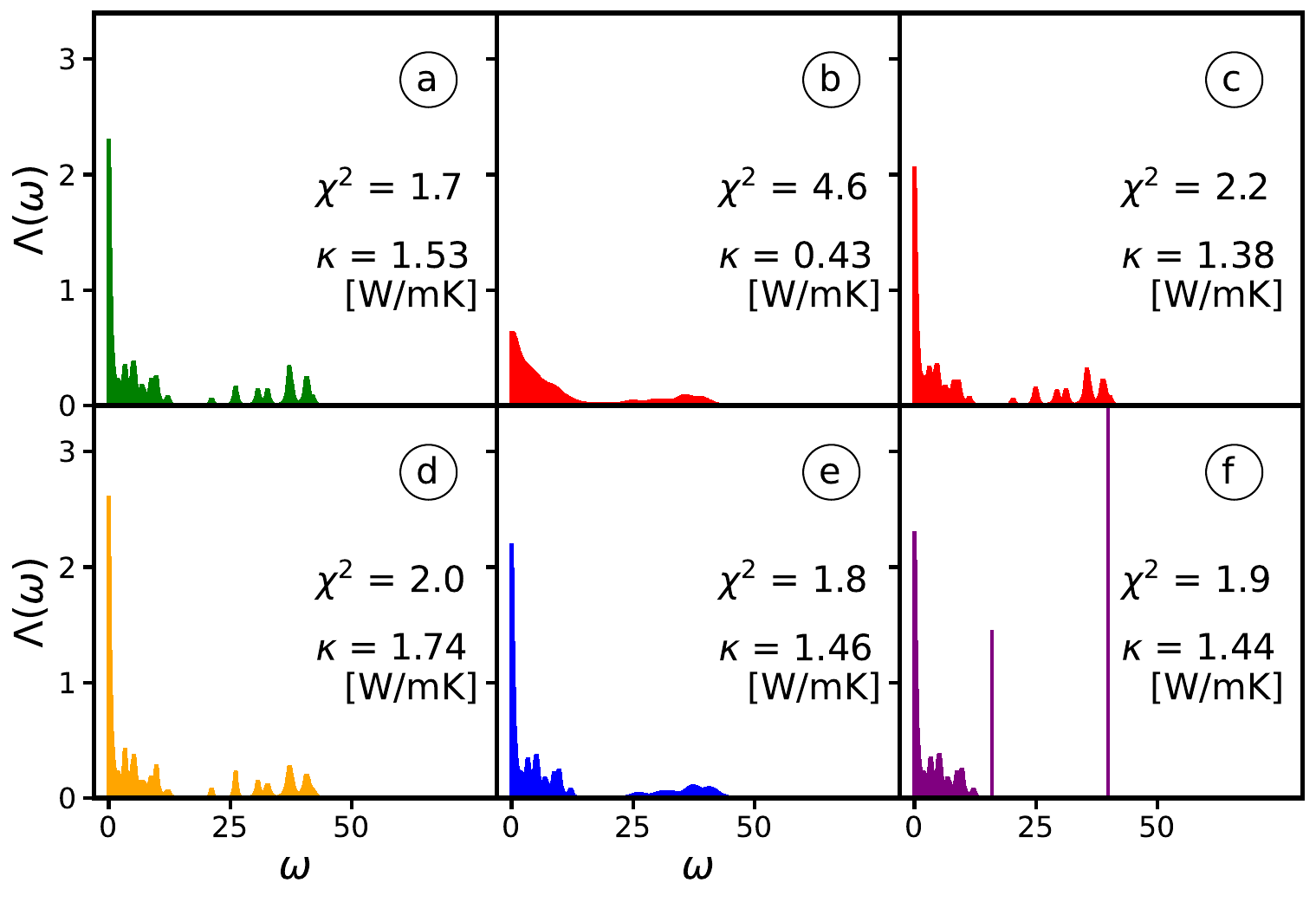}
\caption{Comparison of the spectral reconstructions obtained by using model (a) to (f), given in Eqs~(\ref{ModelA}-\ref{ModelF}) together with the values of $\chi^2$ and the
resulting thermal conductivity.
}
\label{Fig::reconstructions}
\end{figure}

The most important ingredient 
for the reconstruction of the spectral function
is the separation into low and high frequency regions, where
the low frequency part is
determined by  a single parameter (at most).
In the harmonic approximation,
 the spectral function naturally provides such a separation
 of low and high frequency regions
in terms of the singular and regular parts, $\Lambda_s(\omega)$ and $\Lambda_r(\omega)$ (see Eqs (11-13) in the main text).
Both parts of the spectral functions parametrically
depend on a 
set of phonon frequencies $\{\omega_n,\}$ as well as a set of 
damping rates $\{ \Gamma_n \}$. 

The first relevant question concerns the sensitivity of the results to the value of $\Gamma$. A natural choice for the decay rates would be the inverse phonon life times, $\Gamma^{ph}_{\mathbf{k}a}$, obtained from the normal mode analysis. Note, that we have determined the phonon lifetimes for each mode individually, so that we can use the full set of lifetimes in the expression of the spectral function, leaving $\xi$ as the only parameter of the model. As shown in Fig.~\ref{Fig::sup_harm_T15_75} , the best reconstruction based on imposing the phonon lifetimes is clearly incompatible with the PIMC data of the imaginary time correlation function at all temperatures, in contrast to our reference model with two parameters $\Gamma$ and $\xi$. Only at the two highest temperatures, the imaginary time correlations from the phonon model gets closer to the PIMC data, but even there, an offset remains.

In the following we introduce and discuss several models allowing for different parametrizations of the spectral functions. In particular, we want to address the sensitivity and robustness of our results, concerning the choice of phonon frequencies --
$\omega_{\mathbf{k}a}$ from our mode analysis vs the harmonic frequencies $\omega_{\mathbf{k}a}^0$ from the bare harmonic interaction potential --,
the mode dependence of the damping rates $\Gamma_{\mathbf{k}a}^{ph}$, 
and more general variations of the regular (high frequency) part.

Therefore, we introduce the following models
\begin{eqnarray}
(a) & \Lambda_{(a)}(\omega | \Gamma,\xi,\omega_{\mathbf{k}a}) &
= \Lambda_\alpha^s(\omega|\Gamma,\{ \omega_{\mathbf{k}a} \})+\xi \Lambda_\alpha^{r}(\omega|\Gamma,\{ \omega_{\mathbf{k}a} \}) \quad \quad \text{(default/reference model)} 
\label{ModelA}\\
(b) & \Lambda_{(b)}(\omega | \Gamma,\xi,\omega_{\mathbf{k}a}^0) &
= \Lambda_\alpha^s(\omega|\Gamma,\{ \omega_{\mathbf{k}a}^0 \})+\xi \Lambda_\alpha^{r}(\omega|\Gamma,\{ \omega_{\mathbf{k}a}^0 \}) 
\label{ModelB}\\
(c) & \Lambda_{(c)}(\omega | \Gamma,\xi_1,\xi_2,\omega_{\mathbf{k}a}^0) &
= \xi_1 \Lambda_\alpha^s(\omega|\Gamma,\{ \omega_{\mathbf{k}a}^0 \})+\xi_2 \Lambda_\alpha^{r}(\omega|\Gamma,\{ \omega_{\mathbf{k}a}^0 \}) 
\label{ModelC}\\
(d) &\Lambda_{(d)}(\omega | \gamma,\xi,\omega_{\mathbf{k}a}) &
= \Lambda_\alpha^s(\omega|\{ \gamma \Gamma_{\mathbf{k}a}^{ph} \},\{ \omega_{\mathbf{k}a} \})+\xi \Lambda_\alpha^{r}(\omega|\{ \gamma \Gamma_{\mathbf{k}a}^{ph} \},\{ \omega_{\mathbf{k}a} \})
\label{ModelD}\\
(e) &\Lambda_{(e)}(\omega | \Gamma,\xi,\omega_{\mathbf{k}a}) &
= \Lambda_\alpha^s(\omega| \Gamma,\{ \omega_{\mathbf{k}a} \})+\xi \Lambda_\alpha^{r}(\omega|\{ \Gamma_{\mathbf{k}a}^{ph} \},\{ \omega_{\mathbf{k}a} \})
\label{ModelE}\\
(f) &\Lambda_{(f)}(\omega | \Gamma,\xi,\nu_1,\nu_1,\nu_2) &
= \Lambda_\alpha^s(\omega| \Gamma,\{ \omega_{\mathbf{k}a} \})++\xi_1\delta(\omega-\nu_1+ \xi_2 \delta(\omega-\nu_2) \})
\label{ModelF}
\end{eqnarray}

In the following we focus on $T=20$K in order to study the
robustness of our model assumptions and the sensitivity of
the reconstruction to the various parameter sets.
The obtained spectral functions are shown in Fig.~\ref{Fig::reconstructions},
the quality of the various reconstructions are quantified by 
calculating the $\chi^2$ value of the difference to the respective
PIMC data of the imaginary time current-current correlations.

Model $(a)$ is our default model which we have used for all results
shown in the main text, which serves as reference: the model is based on
the phonon frequencies $\omega_{\mathbf{k}a}$ obtained from the PIMC normal mode spectra, and a uniform (mode-independent) damping rate $\Gamma$. Parameter
$\xi$ multiplying the regular part is the simplest way to modify the 
high frequency part of the spectrum.

In  model $(b)$ and $(c)$, we want to test out the influence of the phonon spectrum by 
using $\omega_{\mathbf{k}a}^0$ -- the frequencies of the ideal harmonic crystal
obtained from diagonalizing the hessian of the bare Lennard-Jones interaction
potential -- in the calculation of the singular and regular part, $\Lambda^{s/r}_\alpha(\omega)$.  Model $(b)$
performs considerably worse than
model $(a)$ for the reconstruction with an approximately three times larger $\chi^2$ value. Since the two models only differ by the phonon spectra, the use of the full
phonon frequencies $\omega_{\mathbf{k}a}$ is not only physically motivated, but also clearly supported directly by the reconstruction of the current correlation function (see also corresponding discussion for the specific heat).

The thermal conductivity of 
model $(b)$ turns out to be considerably lower
than our default model. 
In order to correct for the use of the bare phonon frequencies $\omega_{\mathbf{k}a}^0$,
affecting primarily the high energy part of the spectrum, 
the mode independent damping rate got effectively enlarged. 
In model $(c)$ we use two independent 
multiplicative parameters, $\xi_1$ and $\xi_2$,
to decouple better singular and regular part,
both still calculated with the bare harmonic 
frequencies $\omega_{\mathbf{k}a}^0$. The quality of the reconstruction improves in model $(c)$, and, importantly, the thermal conductivity raises again to a value comparable to our default model.  
Model $(b)$ and $(c)$ are the only models based on the bare harmonic frequencies.

In model $(d)$ and $(e)$ we included the possibility of mode dependent damping rates
through the use of the 
full set $\{ \Gamma_{\mathbf{k}a}^{ph} \}$
obtained from the PIMC normal mode analysis.
In model $(d)$ we use an overall multiplicative
parameter $\gamma$ to modify the strength of
the damping rates. In model $(e)$, we use
the full set of damping rates  $\{ \Gamma_{\mathbf{k}a}^{ph} \}$ in the regular part of the spectral function, whereas a
single, mode-independent damping rate is used in the singular part. Both models give reasonable $\chi^2$ values for the reconstruction, the 
resulting thermal conductivity varies by about  $10\%$ compared to the default model $(a)$.

In model $(f)$ we model the regular part
of the spectral function by two delta peaks. This is certainly a too crude model to represent all features of the high frequency part of the spectrum. Nevertheless, the resulting $\chi^2$ value shows that the reconstruction
performs equally well. Importantly, the value of the thermal conductivity remains also similar to the reference. 

These various results, and in particular those of model $(f)$, actually illustrate the difficulties of a full spectral reconstruction.
Different (incompatible) features of the spectrum can lead to very similar imaginary time correlation functions, common to many analytical continuation problems. 
A meaningful determination of the thermal conductivity from the peak at the origin is only possible by
constraining the
singular part of the spectrum.
This was done
by a physically motivated model, 
e.g. the density of states underlying $\Lambda_\alpha^s$. The results
obtained by model (a)-(f) show that resulting
values for the thermal conductivity are robust (up to around $10\%$) to modifications of the phonon frequencies determining the density of states, 
and, importantly, to the
precise modelling of the regular part at higher frequencies. 

Let us note that the resulting $\chi^2$ values of models $(c)-(f)$ agree reasonably with model $(a)$ (and also with each other). Further validation on independent PIMC data on the 
imaginary time correlation function can then be used to provide
a better, unbiased selection among them \cite{HO_PIMC}. We have not attempted such an analysis here, since the corresponding values of $\kappa$ are robust and sufficiently close to
 each other so that our conclusions in the main text are not affected.

\subsection{Reconstruction of the total current-current correlation function}

\begin{figure}[h]
\centering
\includegraphics[width=\textwidth]{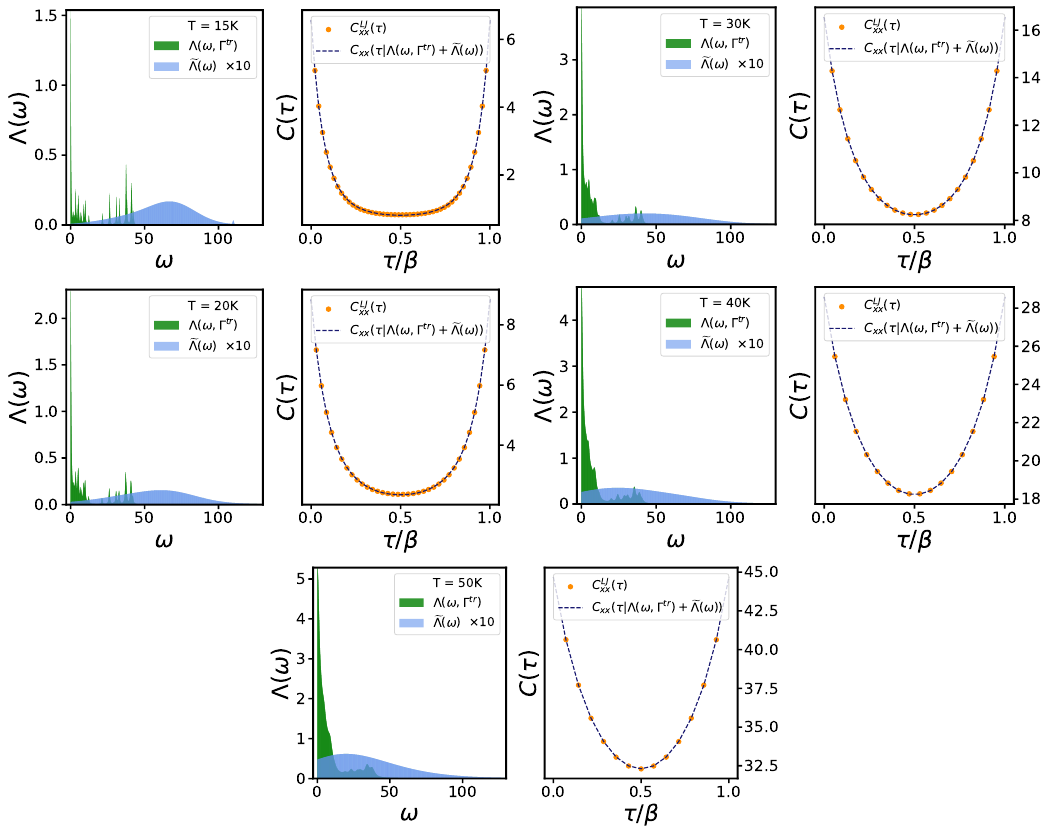}
	\caption{Reconstructed spectral function of total current current correlations at different temperatures, based on the corresponding reconstruction
$\Lambda(\omega,\Gamma^{tr})$ of the harmonic current correlations
and the additional spectral weight $\widetilde{\Lambda}(\omega)$,
as described in the section {\em End Matter} and Fig.~5 of the main text.
Notice the difference in scale between the two components of the spectral density.
The inset shows the reconstructed imaginary time correlation function of the total current obtained by optimizing only the 
additional weight $\widetilde{\Lambda}(\omega)$ 
without modifying $\Lambda(\omega,\Gamma^{tr})$.
}
\label{Fig::Jtot_T15}
\end{figure}

In Fig.~\ref{Fig::Jtot_T15} we show the spectral
functions obtained from the reconstruction of the total current-current correlation functions, as described in the section {\em End Matter}
and shown for $T=20K$ in Fig.~5 of the main text, for all temperatures.

\section{Specific heat}
\label{Sec::specificheat}
\begin{figure}[h]
\centering
\includegraphics[width=0.48\textwidth]{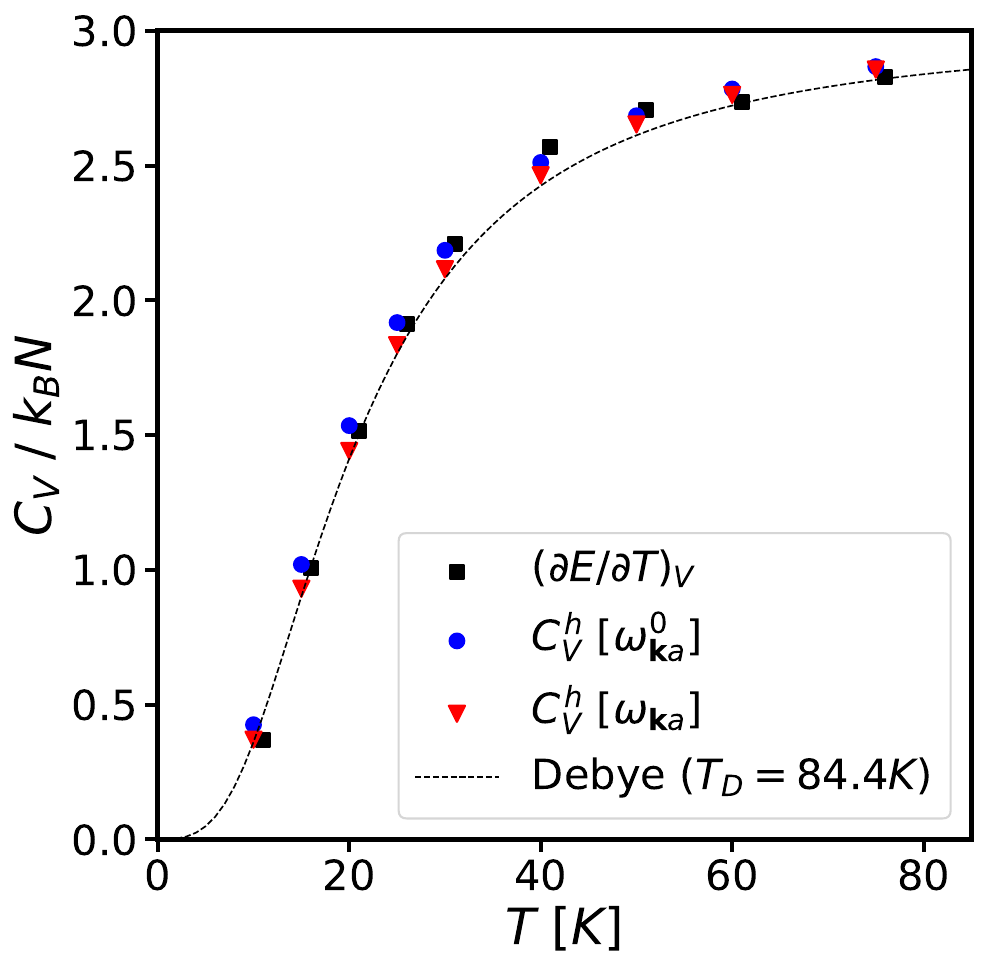}
\caption{Specific heat of solid Ar from PIMC calculations. The black squares are obtained from the numerical derivative of the total energy, whereas $C_V^h$ (triangles) is the specific heat determined in the harmonic approximation based on the phonon frequencies, $\omega_{\mathbf{k}a}$, obtained from the normal mode correlation functions. In addition to Fig.2 in the main text, we also show here the
specific heat in the harmonic approximation, $C_V^0$, using the
bare phonon frequencies, $\omega_{\mathbf{k}a}^0$.
}
\label{Fig::sup_specific_heat}
\end{figure}

In Fig.~\ref{Fig::sup_specific_heat} we also show the improvement
of the specific heat within the harmonic approximation using the
(temperature/density dependent) phonon frequencies, $\omega_{\mathbf{k}a}$,
obtained from the normal mode correlations, compared to the use
of the bare frequencies, $\omega_{\mathbf{k}a}^0$, from the hessian
of the bare (Lennard-Jones) interaction potential.
Whereas at higher temperatures, close to the Debye temperature,
the harmonic approximation -- using either effective phonon or ideal harmonic frequencies -- deviates from the
exact specific heat given by the energy derivative, 
the phonon spectra much better describe the specific heat at lower temperatures. This further supports the use of $\omega_{\mathbf{k}a}$ in our reference model
used for the spectral reconstruction of the heat current.

\section{Phonon and transport lifetimes}

All of our models used for the spectral reconstruction of the 
current correlation function indicates a transport lifetime
$\tau^{tr}$ significantly different from the phonon lifetime.

Instead of a simple mode averaged lifetime,
$\overline{\tau^{ph}}=\sum_{\mathbf{k}a} \tau_{\mathbf{k}a}^{ph}/N$, let us use the following weighted phonon lifetime 
$\overline{\tau^{ph}_{\text{PB-RTA}}}$, defined such that 
\begin{equation}
    \sum_{\mathbf{k}a} c_{\mathbf{k}a}v^2_{\mathbf{k}a} \overline{\tau^{ph}_{\text{PB-RTA}}} =
\sum_{\mathbf{k}a}  c_{\mathbf{k}a}v^2_{\mathbf{k}a}\tau_{\mathbf{k}a}^{ph}
\end{equation}
or
\begin{equation}
     \overline{\tau^{ph}_{\text{PB-RTA}}} =
\frac{\sum_{\mathbf{k}a}  c_{\mathbf{k}a}v^2_{\mathbf{k}a}\tau_{\mathbf{k}a}^{ph}}
{\sum_{\mathbf{k}a} c_{\mathbf{k}a}v^2_{\mathbf{k}a}}
\label{Eq::taumean}
\end{equation}

\begin{figure}[h]
\centering
\includegraphics[width=0.45\textwidth]{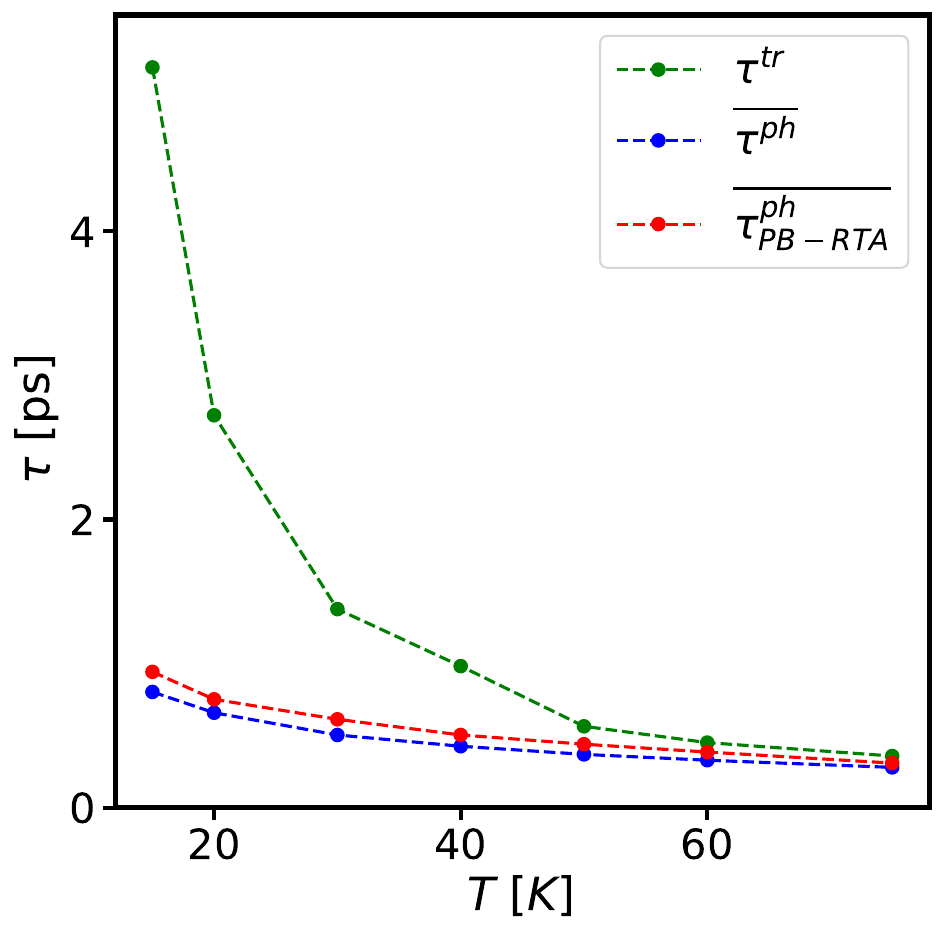}
	\caption{Transport lifetime $\tau^{tr}$ obtained from the
	reconstruction of the current correlation function compared to 
	mean averaged equilibrium lifetime $\overline{\tau^{ph}}$ and to the weighted equilibrium phonon lifetime $\overline{\tau^{ph}_{\text{PB-RTA}}}$,
	defined in Eq.~(\ref{Eq::taumean}).
	\label{Fig::lifetimes}
}
\end{figure}

From Fig.~\ref{Fig::lifetimes}, we see that differences between
$\overline{\tau^{ph}}_{\text{PB-RTA}}$ and $\overline{\tau^{ph}}$ are minor, however the transport lifetime $\tau^{tr}$ obtained from 
our spectral reconstruction increases much faster  lowering the temperature than the averaged or weighted phonon lifetimes.
Roughly, this change of behavior can be understood by noting that
the normal modes (phonons) remain damped even at zero temperature
due to quantum effects, whereas the transport lifetime diverges,
as no energy dissipation can occur in the zero temperature limit.

For completeness, we also show our second parameter $\xi(T)$, multiplying the regular part of the spectrum in our default model, as a function of temperature. Barring broadening, the function $\Lambda_\alpha^r(\omega)$ 
of the regular parts represents the weights of all possible phonon absorption and emission processes occurring within purely harmonic dynamics. Anharmonicity effects in the dymamics will in general not only lead to broadening and renormalization of the phonon frequencies, but also enable higher order processes, absent in the harmonic approximation. The behavior of $\xi(T)$, starting from $\approx 1$ when extrapolated to zero temperature, and growing with $T$, shows the growing importance of beyond harmonic dynamics
at higher temperatures. 

\begin{figure}[h]
\centering
\includegraphics[width=0.45\textwidth]{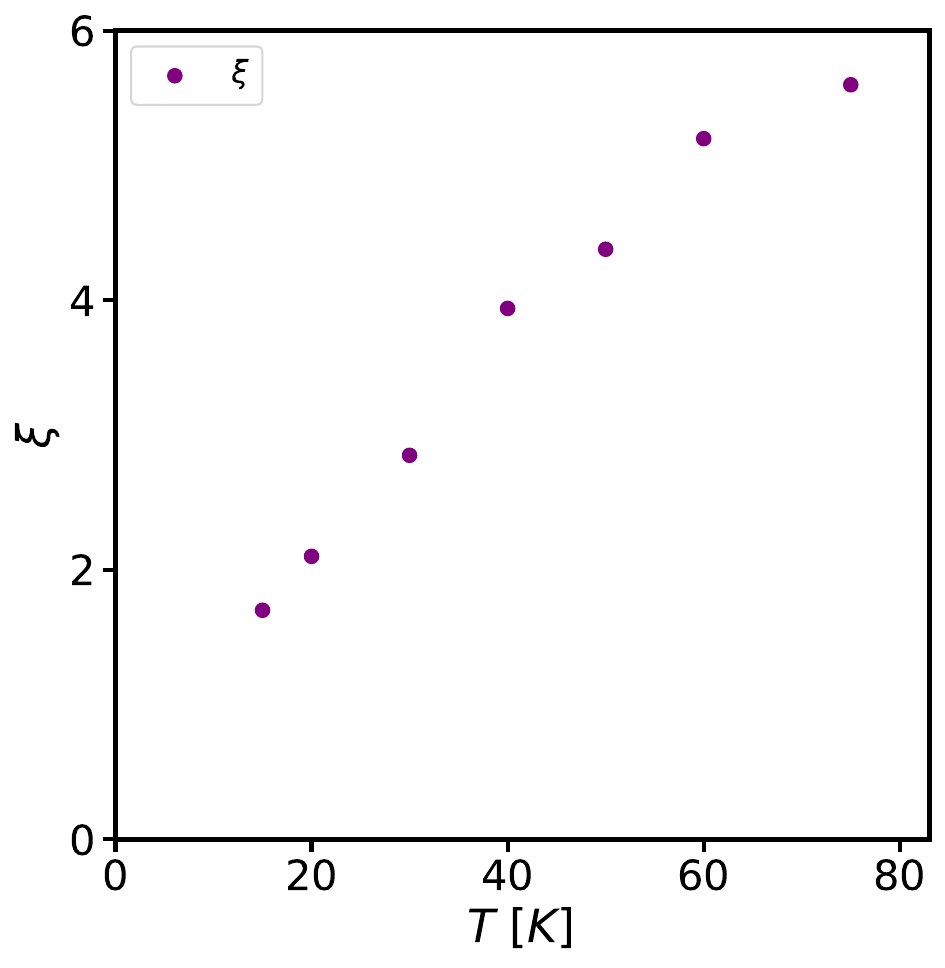}
	\caption{Plot of the parameter $\xi(T)$, multiplting the regular part of the spectrum in our default model, as a function of temperature.
}
\label{Fig::xiT}
\end{figure}

\newpage

\section{Comparison with approximate calculations
of the heat conductivity of solid Argon}

In the main text, we focused on the comparison between the
thermal conductivity obtained from the spectral reconstructions of the imaginary time current-current correlation functions
with respect to the use of the simple Peierls-Boltzmann 
formula using the phonon life times of the normal modes.
We attributed the difference to a modified transport lifetime.

\begin{figure}[h]
\centering
\includegraphics[width=0.45\textwidth]{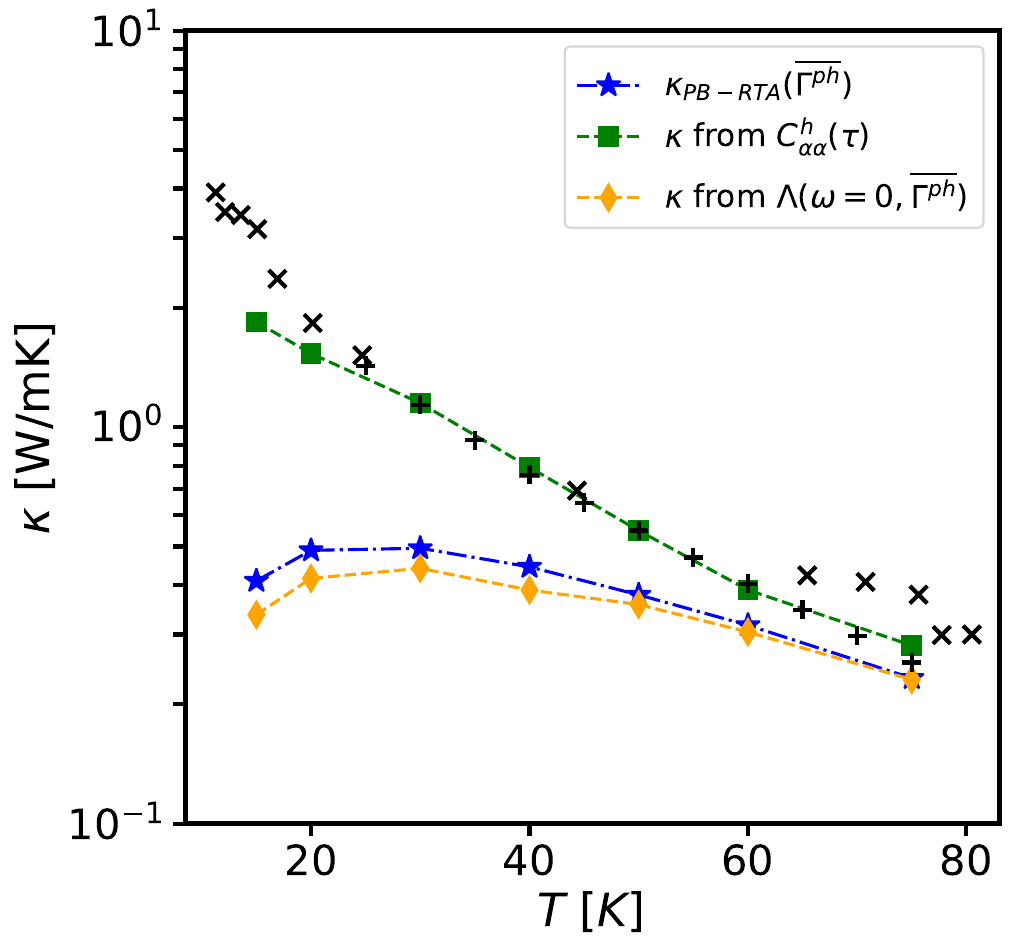}
\includegraphics[width=0.45\textwidth]{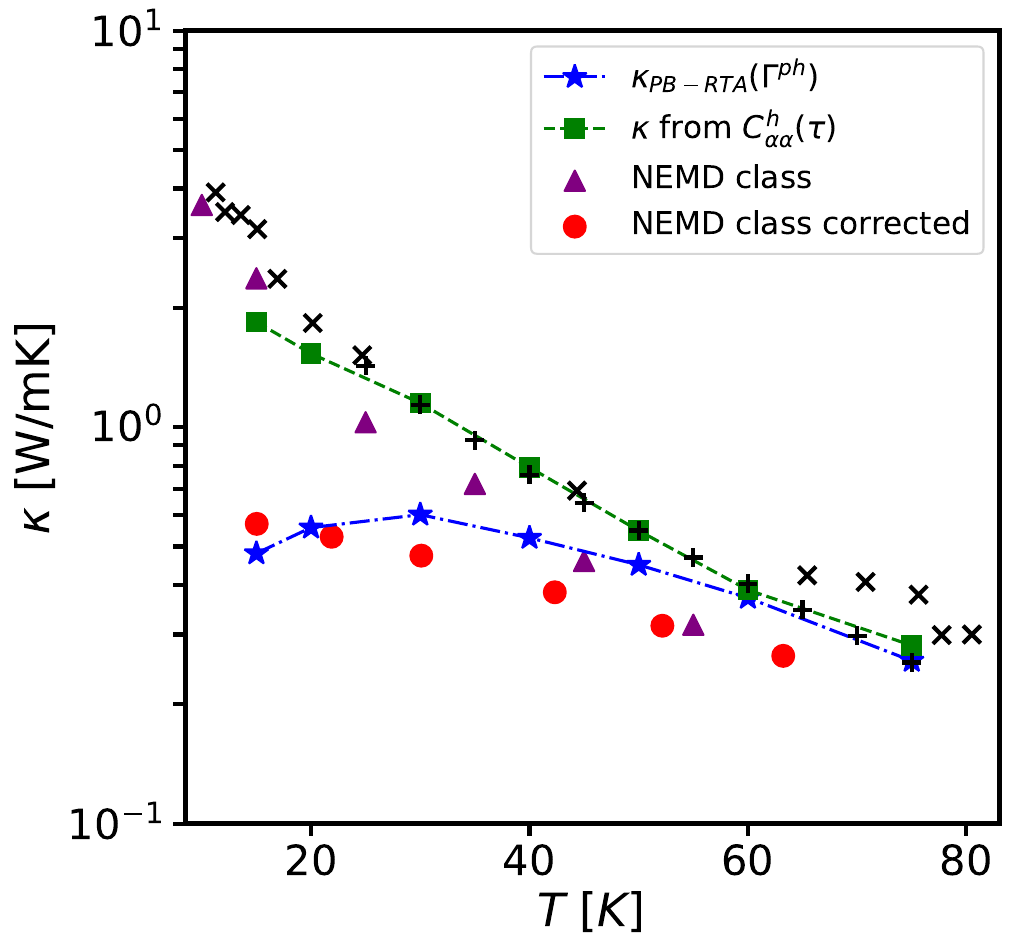}
	\caption{On the left, we compare the thermal conductivity of Argon with
	the results from the Peierls-Boltzmann formula using the equilibrium phonon lifetimes determined by the normal modes, $\kappa_{\text{PB-RTA}} = \sum_{\mathbf{k}a} c_{\mathbf{k}a}v^2_{\mathbf{k}a}\tau_{\mathbf{k}a}^{ph}$ and the corresponding results from the quasi-harmonic spectral function $\Lambda(\omega | \{ \Gamma_{\mathbf{k}a}^{ph} \}, \{\omega_{\mathbf{k}a}^{ph} \})$. Here, the quasi-harmonic spectral function is calculated using the same (normal mode) phonon frequencies and liftimes calculated at thermal equilibrium
	as in the Peierls-Boltzmann formula, so the main differences
	amounts to the inclusion of off-diagonal matrix elements of the velocity matrix. Both approximations, essentially relying on the phonon lifetimes
	do not reproduce the strong rise of the conductivity at low temperatures,
	observed in experiments and reproduced by our reference reconstructions based on the
	transport lifetime. On the right figure, we show the results of classical and quantum corrected NEMD calculations of Ref.~\cite{JLB2014}.
}
\label{Fig::kappa_NEMD_others}
\end{figure}

Let us first study the influence of small modifications of the Peierls-Boltzmann formula. First, one might replace the set of
phonon lifetimes 
by the simple mode averaged lifetime. The resulting changes
are directly given by the changes from $\overline{\tau^{ph}_{\text{PB-RTA}}}$ to $\overline{\tau^{ph}}$, which 
are minor (see Fig.~\ref{Fig::lifetimes}).
Similarly, we have also computed the thermal conductivity
corresponding to $\Lambda^s_\alpha(\omega=0|\{ \Gamma_{\mathbf{k}a}^{ph} \}, \{ \omega_{\mathbf{k}a}^{ph} \})$ from Eq.(12) of the main text, using the
phonon lifetimes and frequencies. This corresponds to the use of the full velocity matrix instead of diagonal approximation 
used in the Peierls-Boltzmann formula, so taking into account interband transport. As we see from Fig.~\ref{Fig::kappa_NEMD_others}, the effect of off-diagonal
matrix elements are also minor, and cannot explain the 
strong rise lowering temperature.
This supports our physical interpretation that low temperature
rise is essentially due to a sizeable modification of the
lifetime themselves.

In the following, we compare our results with different approximate calculations. First, we show previous results from  Ref.~\cite{JLB2014}
based  on classical and quantum corrected NEMD calculations.
Although in rough agreement with experiments,
the strong rise at low temperatures observed in the classical NEMD results 
must be considered as an artifact of the purely classical approach, 
as it qualitatively changes when quantum corrections, needed to
reproduce the correct specific heat, are included. Classical NEMD results 
corrected for quantum effects are closer to our Peierls-Boltzmann results.

\begin{figure}[h]
\centering
\includegraphics[width=0.48\textwidth]{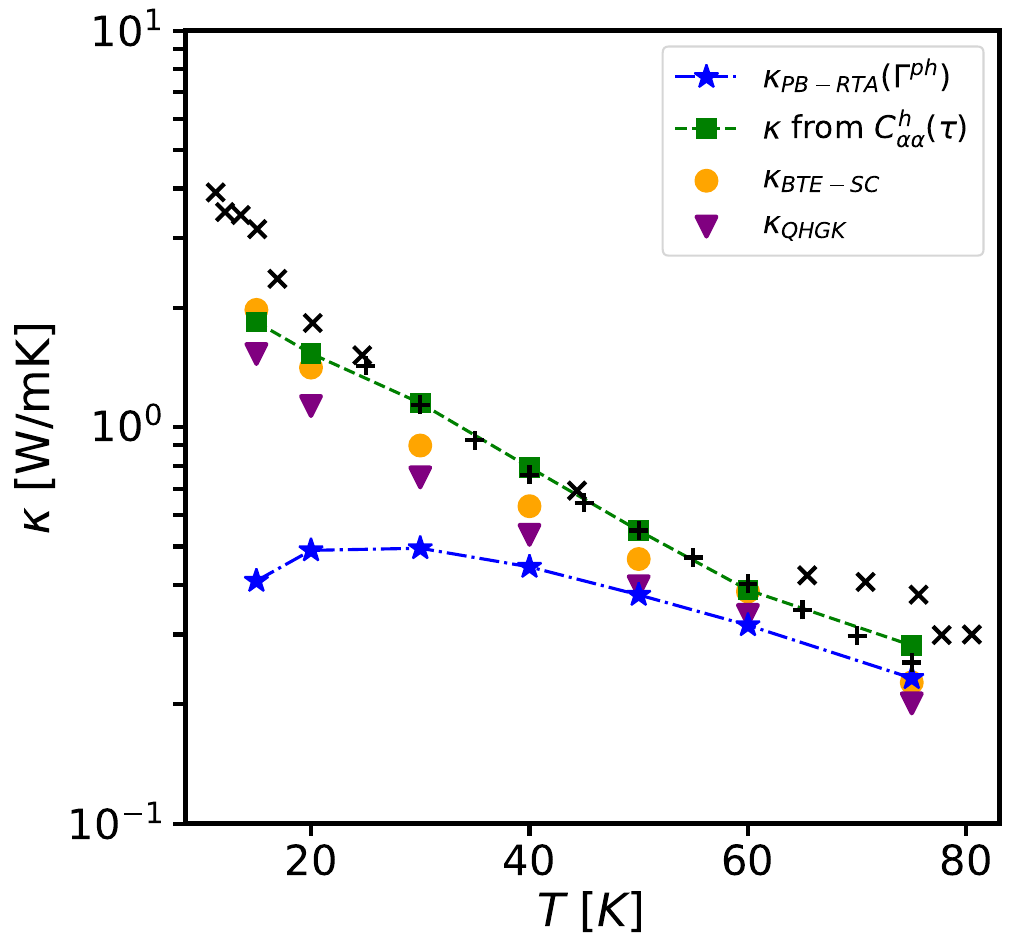}
	\caption{Comparison of the thermal conductivity of Argon with results from QHGK and BTE (self-consistent iteration) calculations computed using the implementation in \cite{kaldo,Baroni2019,KaldoWeb}.
}
\label{Fig::kappa_qhgk}
\end{figure}

In Fig.~\ref{Fig::kappa_qhgk}, we compare with a
more recent implementation of the quasi-harmonic 
Green-Kubo approximation (QHGK) and the Boltzmann transport equation (BTE)
KALDO Ref.~\cite{kaldo,Baroni2019,KaldoWeb}, including cubic effects in the anharmonicity.
Similar to the purely classical NEMD predictions, both approaches provide a much better description than the quantum corrected NEMD.
However, as pointed out in Ref.~\cite{Feng2016}, quartic effects of the anharmonicity (4 phonon scattering rates) are important for Argon, reducing the thermal conductivity by up to $60\%$ compared to cubic approximation. The reasonable 
agreement of QHGK and BTE calculations with experiment indicate
possible error cancellations in the calculation of solid Argon,
by the cubic approximation together with the neglect of vertex corrections.
In fact, as discussed in the following section, both approaches, BTE and QHGK, do not perform
as good
for solid Neon.


\newpage

\section{Thermal Conductivity of solid Neon}

\begin{figure}[h]
\centering
\includegraphics[width=0.48\textwidth]{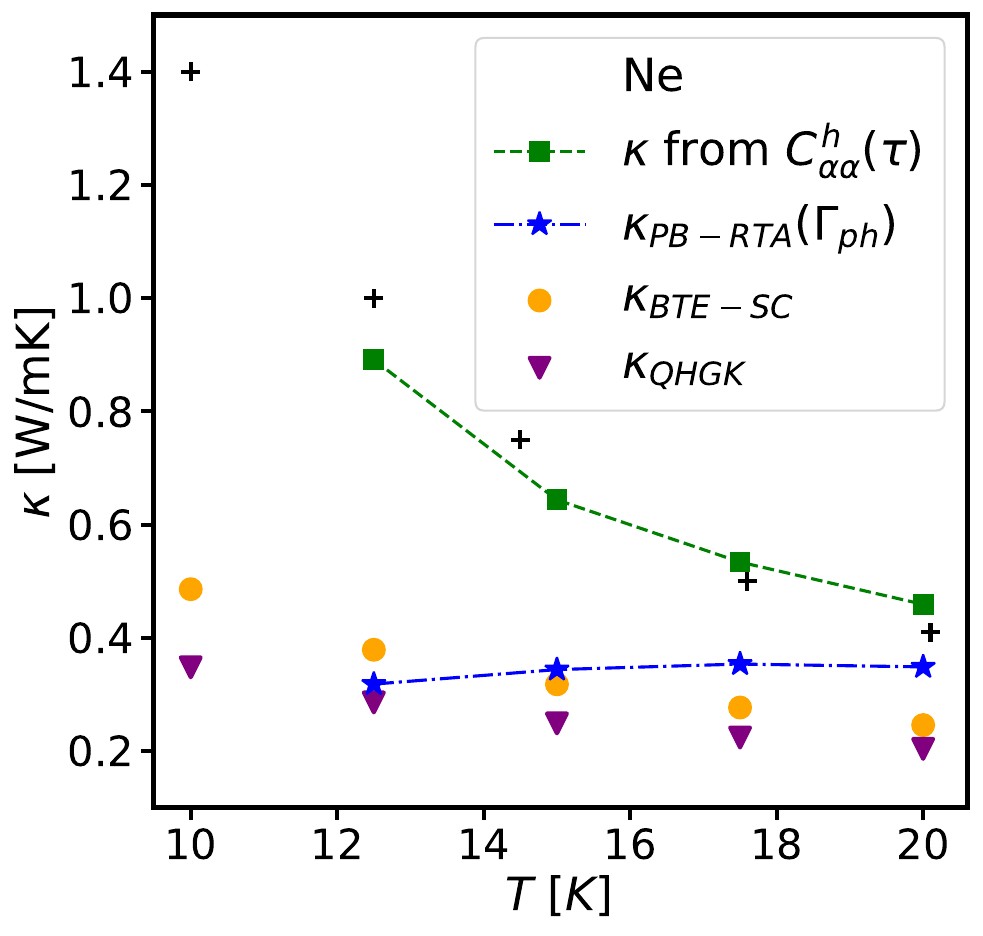}
\caption{Thermal conductivity of solid Neon obtained from the spectral reconstruction of $C_{xx}(\tau)$, compared with the result of the Peierls-Boltzmann equation, $\kappa_{\text{PB-RTA}}$ based on the PIMC phonon frequencies and lifetimes. The experimental data from \cite{Weston1984} at corresponding crystal density is indicated as $+$. QHGK and BTE (self-consistent iteration) calculations computed using the implementation in \cite{kaldo,Baroni2019,KaldoWeb} based on three phonon scattering rates anharmonicity are shown for comparison.
}
\label{Fig::kappa_current_Ne}
\end{figure}

The article primarily focused on the description of solid Argon, 
modelled by a Lennard-Jones potential. As an independent verification
of our methods, we have also applied our methodology to calculate
the heat conductivity of solid Neon, shown in Fig.~\ref{Fig::kappa_current_Ne},
using $\epsilon=36.68$K, and $\sigma=2.787$\AA \cite{Cuccoli1993}. The particle density is kept fixed $\rho \sigma^3 \simeq 0.965$ which corresponds approximately to the zero pressure isobar. Due to the smaller mass of the Neon atoms, the dimensionless parameter $Q=\sqrt{\hbar^2/m \sigma^2 \epsilon} \simeq .0918$ is roughly three times bigger than the one of Argon. 
The Debye temperature of Neon is $68.5$K. 
Calculations have been performed with an imaginary time step corresponding to $\approx 700$K.
We have applied the same methodology for solid Neon. In Fig.~\ref{Fig::kappa_current_Ne} we show our PIMC results for the thermal conductivity form the reconstruction of the imaginary time correlations of the harmonic current determining the a transport lifetime, and the corresponding results of the
Bethe-Peierls formula using the phonon lifetimes. Again, the values obtained from the reconstruction are increasing at lower temperatures, in reasonable close agreement to experiment \cite{Weston1984}.
For comparison, we also show the results of QHGK and BTE using the implementation of Ref.~\cite{kaldo,Baroni2019,KaldoWeb}.

\end{document}